\documentclass[twocolumn,preprint]{aastex62}
\usepackage[T1]{fontenc}
\usepackage[english]{babel}
\usepackage{natbib}
\usepackage{ulem,xcolor}
\usepackage{amsmath, graphicx}
\usepackage{soul}
\usepackage{lineno}
\def\rhessi{{\textit{RHESSI}}}
\def\fermi{{\textit{Fermi}}}
\def\goes{{\textit{GOES}}}

\def\kw{{Konus-\textit{Wind}}}
\def\sdo{{\textit{SDO}}}

\def\mw{{microwave}}

\def\MW{{MICROWAVE}}

\def\gs{{gyrosynchrotron}}

\begin{document}

\title{Energy budget in the 2017-09-07 ``cold'' solar flare}

	\author[0000-0001-5557-2100]{Gregory D. Fleishman}
	\affil{Center For Solar-Terrestrial Research, New Jersey Institute of Technology, Newark, NJ 07102, USA}

\affil{Institut f\"ur Sonnenphysik (KIS), Sch\"oneckstrasse 6, D-79104 Freiburg, Germany}

\author[0000-0001-7856-084X]{Galina G. Motorina}
	\affil{Central  Astronomical Observatory at Pulkovo of Russian Academy of Sciences, St. Petersburg, 196140, Russia}


\author[0000-0003-2872-2614]{Sijie Yu} 
	\affil{Center For Solar-Terrestrial Research, New Jersey Institute of Technology, Newark, NJ 07102, USA}

\author[0000-0003-2846-2453]{Gelu M. Nita}
	\affil{Center For Solar-Terrestrial Research, New Jersey Institute of Technology, Newark, NJ 07102, USA}


\begin{abstract}
A subclass of early impulsive solar flares, cold flares, was proposed to represent a clean case, where the release of the free magnetic energy (almost) entirely goes to acceleration of the nonthermal electrons, while the observed thermal response is entirely driven by the nonthermal energy deposition to the ambient plasma. This paper studies one more example of a cold flare, which was observed by a unique combination of instruments. In particular, this is the first cold flare observed with the Expanded Owens Valley Solar Array and, thus, for which the dynamical measurement of the coronal magnetic field and other parameters at the flare site is possible. With these new data, we quantified the coronal magnetic field at the flare site, but did not find statistically significant variations of the magnetic field within the measurement uncertainties. We estimated that the uncertainty in the corresponding magnetic energy exceeds the thermal and nonthermal energies by an order of magnitude; thus, there should be sufficient free energy to drive the flare. We discovered a very prominent soft-hard-soft spectral evolution of the \mw-producing nonthermal electrons. We computed energy partitions and concluded that the nonthermal energy deposition is likely sufficient to drive the flare thermal response similarly to other cold flares.
\end{abstract}



\section{Introduction}
\label{S_Intro}
Solar flares are commonly referred to {as} transient brightenings in the solar atmosphere driven by release of free magnetic energy \citep{2008LRSP....5....1B}.
A typical solar flare demonstrates substantial spatial complexity and time variability in both thermal and nonthermal components. This includes multiple episodes of nonthermal particle acceleration and plasma heating---both associated \citep[Neupert effect, ][]{1968ApJ...153L..59N} and not associated \citep[direct heating; see, e.g.,][]{2010ApJ...725L.161C} with the nonthermal energy deposition. The chain of the magnetic energy conversions to other types of energy such as thermal, nonthermal, and kinetic energies and transformations between them is extremely complex. 
The flare complexity, spatial and temporal, further complicates understanding of these phenomena and their quantification. Thus, analysis of flares with simple time profiles, for example,  single rise-decay shape (single-spike flares), may represent a cleaner case for the flare energy release and budget study compared with other cases. 

Often, although not always, the single-spike flares belong to a class of early impulsive flares \citep{2006ApJ...645L.157S} or even to an extreme subclass of ``cold'' flares \citep{2018ApJ...856..111L, 2023ApJ...954..122L}. 
\citet{2018ApJ...856..111L} proposed that the cold flares may represent a cleanest case for the study of particle acceleration and their forthcoming thermalization resulting in a thermal response of the ambient plasma.
However, several case studies of the cold flares \citep{Fl_etal_2016,2020ApJ...890...75M} revealed that the spatial structure of the flaring volume was not that simple---it consisted of two distinct (likely, interacting) magnetic flux tubes. In other cases \citep[e.g.,][]{2021ApJ...913...97F}, more than two flaring loops may be involved. 
Perhaps, having more than one flux tube is the morphology that facilitates the release of the free magnetic energy to drive the flare. 

Quantification of the energy release and forthcoming transformation requires remote diagnostics of the magnetic, thermal, and nonthermal energies (and, in a general case, the kinetic energy as well). While the quantification of the thermal and nonthermal energies has been available from the extreme ultraviolet (EUV) and soft X-ray (SHR) measurements in the former case and from microwave and hard X-ray (HXR) measurements in the latter case, the magnetic energy has been estimated only indirectly---based on coronal extrapolations of the magnetic field measured at the photosphere.

Recently, a new powerful methodology capable of measuring the evolving magnetic field in the coronal flaring loops has been reported \citep{2020Sci...367..278F}, thus, offering a new avenue of estimating the magnetic energy in the flaring volume. This methodology is based on the broadband microwave imaging spectroscopy now available from the Expanded Owens Valley Solar Array \citep[EOVSA,][]{Gary_etal_2018} and, along with the spatially resolved measurements of the magnetic field 
\citep{2020Sci...367..278F}, also offers maps of the thermal plasma number density and nonthermal electron population \citep{2022Natur.606..674F}.

\begin{figure}\centering
\includegraphics[width=0.99\columnwidth]
{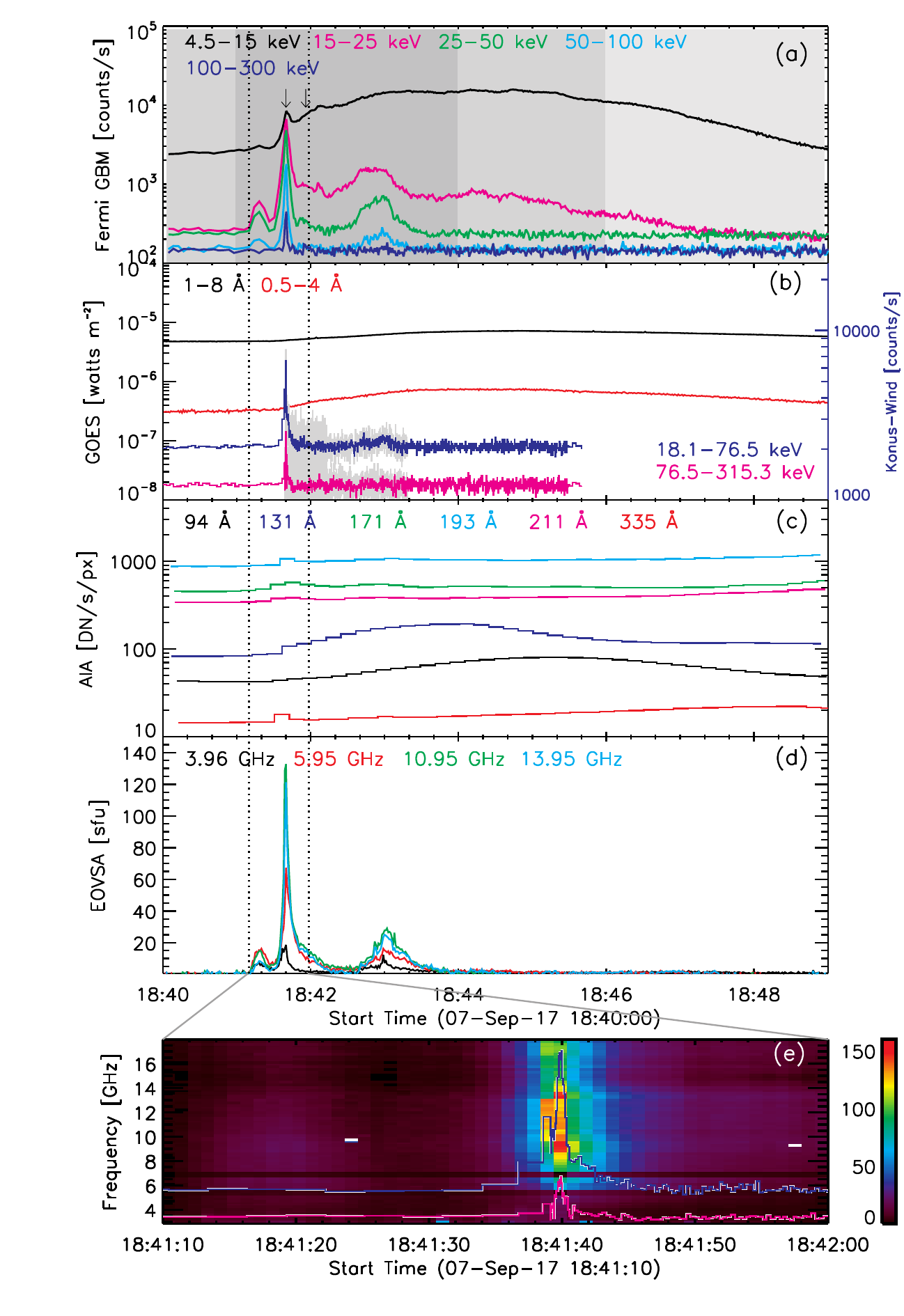}
\caption{Overview of the 2017-Sep-07 flare. From top to bottom: (a) \fermi/GBM light curves at several energy ranges indicated by the legend. The gray---dark gray---gray---light gray  areas indicate the breakdown onto the time ranges, where the model spectral fitting of the \fermi/GBM data was performed using averaging over, respectively, 20---4---20---60\,s time intervals. Vertical arrows point on time frames, {where the \fermi/GBM} spectra and fits are shown in Figure \ref{Fig_fermi_fits} {(middle and right panels)}; (b) low and high \goes\ channels and G1 and G2 \kw\ light curves. The full resolution \kw\ light curves are shown in light gray. To reduce the statistical fluctuations the light curves were averaged to 256\,ms after the burst, observed with 16\,ms time cadence. The averaged light curves are shown in blue and scarlet; 
(c) \sdo/AIA light curves, for six passbands indicated in the panel, obtained from the selected ROI (see Figure \ref{Fig_aiaFOV}); {(d) EOVSA light curves at several microwave frequencies indicated with the legend;} (e) EOVSA dynamic spectrum of the impulsive flare phase with overlaid \kw\  light curves plotted in arbitrary units. Vertical dotted lines in panels (a-d) indicate the impulsive phase shown in the (e) panel. 
\label{Fig_lightcurves}
}
\end{figure}

Here we study a solar flare recorded by EOVSA on 2017-Sep-07 around 18:41 UT and observed by a handful of other instruments. {This flare demonstrates a single main nonthermal peak, although two or three weaker peaks are also present, and formally falls into a category of ``cold'' flares according to the classification proposed by \citet{2023ApJ...954..122L} based on the relationship between the  X-ray and \mw\ fluxes; see Figure\,\ref{Fig_lightcurves}.}
We report on multi-instrument observations of this flare, quantification of its thermal and nonthermal energies with the EUV, SXR, and HXR diagnostics, physical parameter maps derived from the microwave diagnostics, and devise a 3D model of the flare.

\section{Observations} \label{S_Observations}

The solar flare SOL2017-09-07T184140, GOES class C4.5, occurred at $\sim$18:41~UT in AR 12673 with $\beta\gamma$-configuration located at W770S208 just after a C5.2 class flare which ended at 18:36~UT in the same AR.  The flare displayed {several peaks, the main of which had} a short impulsive profile with a peak at $\sim$18:41:40\,UT in hard X-ray (HXR) above 20\,keV and in \mw s that lasted about 20\,s. This short impulsive event shares some properties of nonthermally-dominated  (`cold') solar flares \citep{2020ApJ...890...75M} and it is the first of them for which microwave imaging spectroscopy data from EOVSA were available.



\subsection{Overview of the instruments used in the analysis and gaps in the data}

\begin{figure*}\centering
\includegraphics[width=16cm]{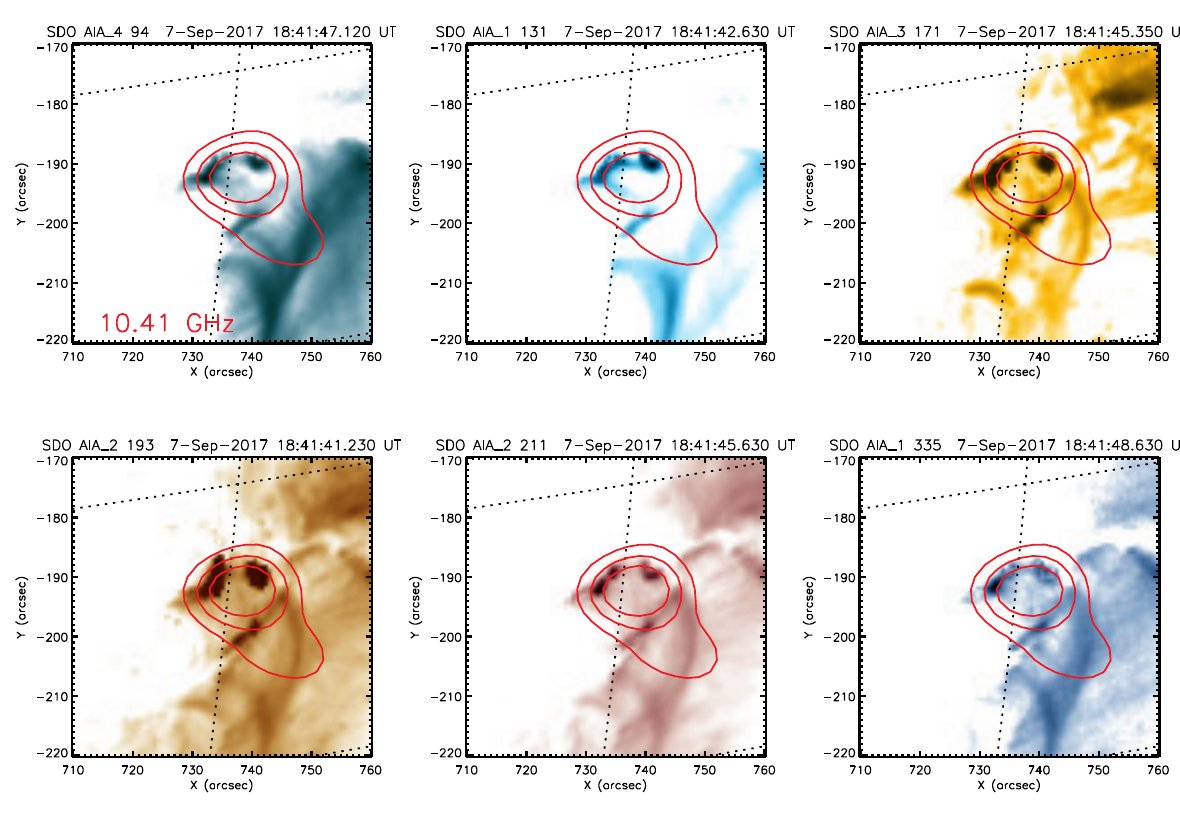}
\caption{{AIA maps taken during the impulsive phase of the September 07, 2017 flare  with overlaid EOVSA map 30, 50, and 70\% contours at 10.41 GHz (red lines) taken at 18:41:39.5~UT, which corresponds to the
maximum of the microwave emission.}    
\label{Fig_aiaFOV}
}
\end{figure*}

This flare was observed with a unique combination of space- and ground-based instruments throughout the entire electromagnetic spectrum, from radio waves to X-rays, see Figure~\ref{Fig_lightcurves}. The spectral X-ray data are available from the {Gamma-ray  Burst Monitor (GBM) on board of the Fermi Gamma-ray Space Telescope  (\fermi/GBM; \citealt{2009ApJ...702..791M}) and \kw\ \citep{1995SSRv...71..265A}, while \rhessi\ \citep{2002SoPh..210....3L} } data are only available in the late decay phase, and are not too useful in this case. {The \fermi/GBM and \kw\ light curves are shown in Figure~\ref{Fig_lightcurves}(a) and  Figure~\ref{Fig_lightcurves}(b), respectively.} Contextual soft X-ray {data from the Geostationary Operational
Environmental Satellite (GOES; \citealt{2005SoPh..227..231W})} are available (Figure~\ref{Fig_lightcurves}(b)). In the EUV domain, {\textit{Solar Dynamics Observatory}/Atmospheric Imaging Assembly (\sdo/AIA; \citealt{2012SoPh..275...17L})} data are available to quantify the thermal properties of the flaring plasma. {The EUV light curves integrated over the region of interest (ROI; see Figure~\ref{Fig_aiaFOV}) are shown in Figure~\ref{Fig_lightcurves}(c).} Unique data, not available for the cold solar flares studied before, but available in our case, are the microwave imaging spectroscopy data from EOVSA and, thus, the corresponding data products such as evolving parameter maps \citep{2020Sci...367..278F,2022Natur.606..674F}. {The EOVSA {light curves and the dynamic spectrum are} shown in Figure~\ref{Fig_lightcurves}(d, e).} The data presented in Figure~\ref{Fig_lightcurves} permit us to estimate the increase of the GOES flux, $\Delta GOES\simeq  4.9\times10^{-7}$ W\,m$^{-2}$, during the flare impulsive phase, and to compare it with the microwave peak flux of 150\,sfu at 9.4~GHz, which confidently places this flare in the category of the cold ones; cf. figure\,11 in \citet{2023ApJ...954..122L}.

\subsection{X-ray data}


{Hard X-ray observations of the impulsive phase of the flare are only accessible from \fermi/GBM and \kw. The \fermi/GBM light curves (see Figure~\ref{Fig_lightcurves}(a)) show {several} peaks, where the second one at ~$\sim$18:41:39.7~UT is associated with the main peak of the microwave emission.}

{\kw\  recorded the impulsive phase of the flare in the G1 and G2 channels both in the waiting and trigger modes, where  its trigger time at the Earth center was 18:41:39.64~UT. The \kw\ count rate (see Figure~\ref{Fig_lightcurves}(b)) 
 had quite high background level during the flare. Thus, \kw\ didn't see the weaker peak seen by \fermi/GBM. The HXR peak time seen by \kw\ in the 18.1-76.5 keV range was at 18:41:39.9~UT, which will be referred as the HXR peak time further in the text. 
 }
 

{\goes\ soft X-ray (SXR) data are shown in Figure\,\ref{Fig_lightcurves}(b).  
There is slight enhancement seen in both low and high energy channels at $\sim$18:41\,UT, which is roughly co-temporal with  \sdo/AIA enhancement seen in the time profiles in Figure\,\ref{Fig_lightcurves}(c).}

\subsection{EUV: \sdo/AIA data} \label{S_Observations_aia}

\begin{figure*}\centering
\includegraphics[width=0.32\textwidth]{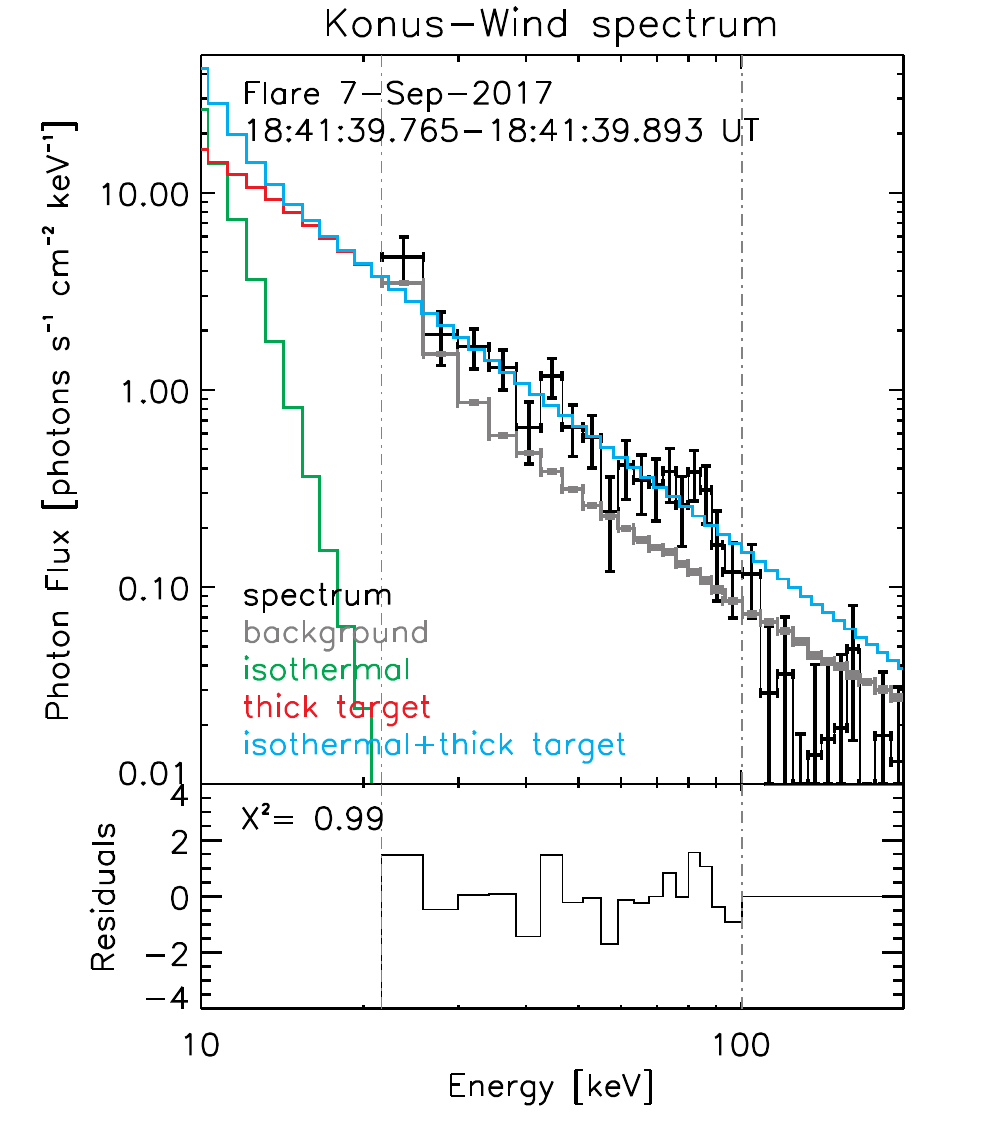}
\includegraphics[width=0.32\textwidth]{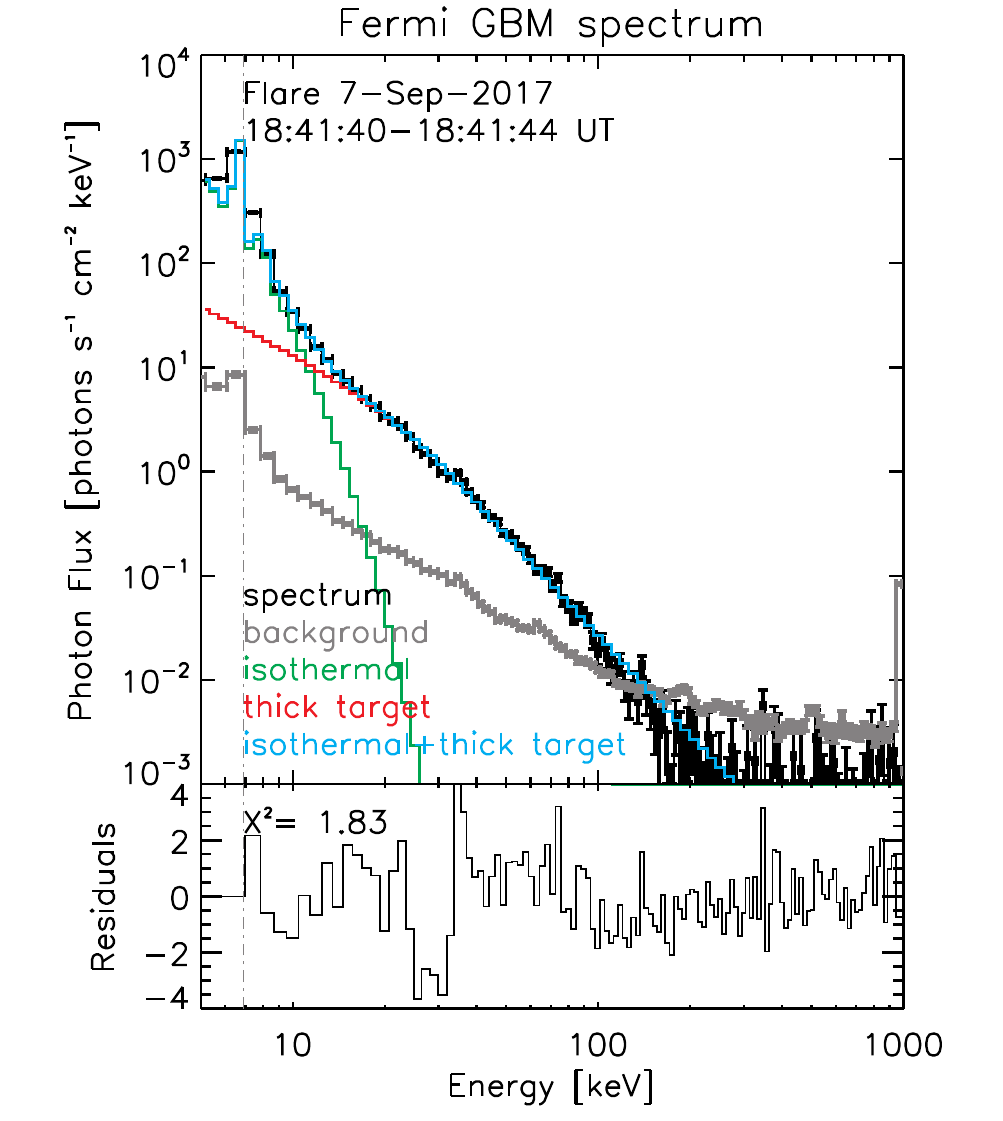}
\includegraphics[width=0.32\textwidth]{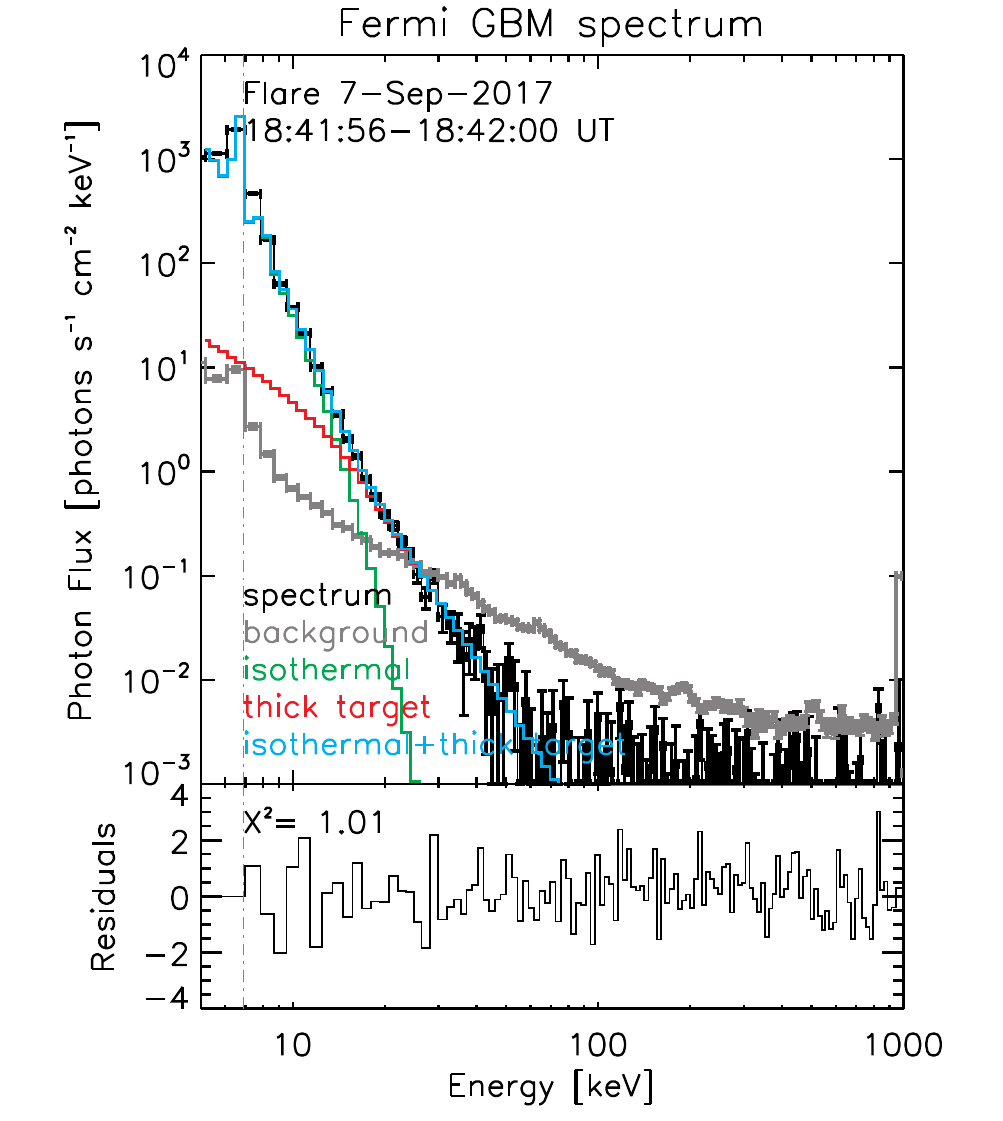}
\caption{{Examples of the \kw\ and \fermi/GBM fits 
(blue histogram) with isothermal (green histogram) plus thick target model (red histogram) during the nonthermal peak at 18:41:39~UT (left panel) and later 18:41:40-18:41:44~UT, 18:41:56-18:42:00~UT (middle and right panels). The X-ray averaged data and the background level are shown with black and gray histograms respectively. Bottom panels indicate residuals.}
\label{Fig_fermi_fits}
}
\end{figure*}

{The EUV full solar disk data set is available in six \sdo/AIA (94, 131, 171, 193, 211, 335 \AA) coronal passbands. The AIA images with high $\sim$1.2" spatial and 12 s temporal resolution have been calibrated to level 1.5 using the \texttt{aia\_prep} routine in SSWIDL and normalized by the exposure time. There was almost no saturation of the AIA data during the flare. The saturation effects are included in further analysis of the AIA data.
We investigate the EUV emission from the ROI shown in Figure~\ref{Fig_aiaFOV} to estimate the thermal energy of the flare plasma and its evolution in the coronal part. 
To infer the integral plasma properties, we use the Differential Emission Measure (DEM) technique \citep{2012A&A...539A.146H} applied to the entire ROI. To derive spatially resolved plasma parameters (temperature and emission measure), we employ the DEM maps \citep{2013A&A...553A..10H} with the methodology developed by \citet{2020ApJ...890...75M} and earlier applied to two solar flares:  SOL2013-11-05T035054 and  SOL2014-02-16T064600 \citep{2020ApJ...890...75M,2021ApJ...913...97F}. 
}

\subsection{MW data} 

This event has been imaged using EOVSA at 134 frequencies, across 31 spectral windows from 3 to 18 GHz, following the methods used by \citet{2020Sci...367..278F}, but with 1-second cadence. The total power dynamic spectrum is shown in Figure~\ref{Fig_lightcurves}. The nominal full-width-half-maximum (FWHM) spatial resolution of this observation is elliptical, with a major axis of $180''/f_{\text{GHz}}$ and a minor axis of $45''/f_{\text{GHz}}$. During the CLEAN imaging process, a circular restoring beam of FWHM $112''.5/f_{\text{GHz}}$ was used. This corresponds to a beam size decreasing from $37''.5$ at 3 GHz to $6''.3$ at 18 GHz.


\section{diagnostics of thermal plasma and non-thermal electrons derived from HXR observations}

\begin{figure*}\centering
\includegraphics[width=0.48\textwidth]{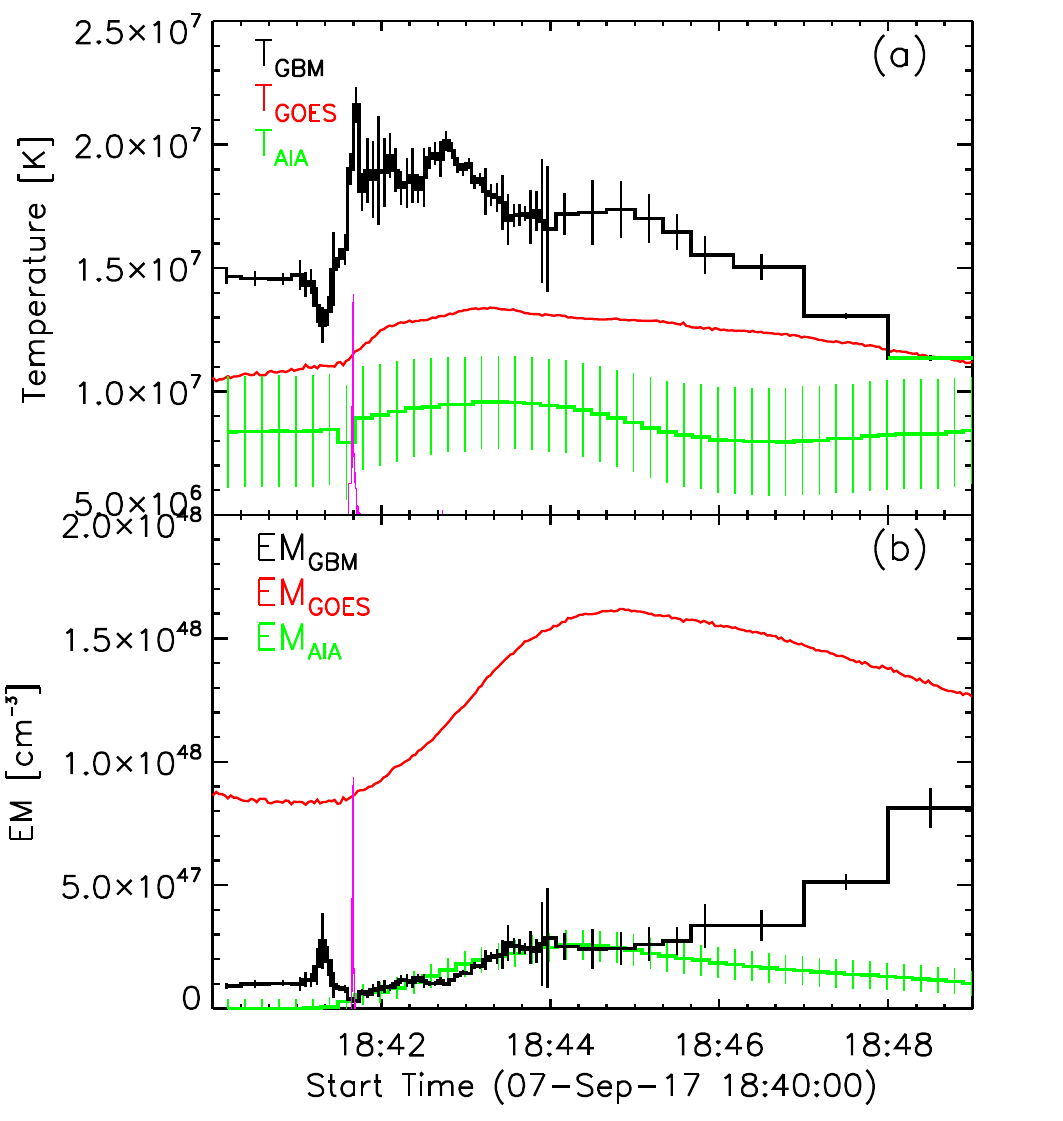}
\includegraphics[width=0.48\textwidth]{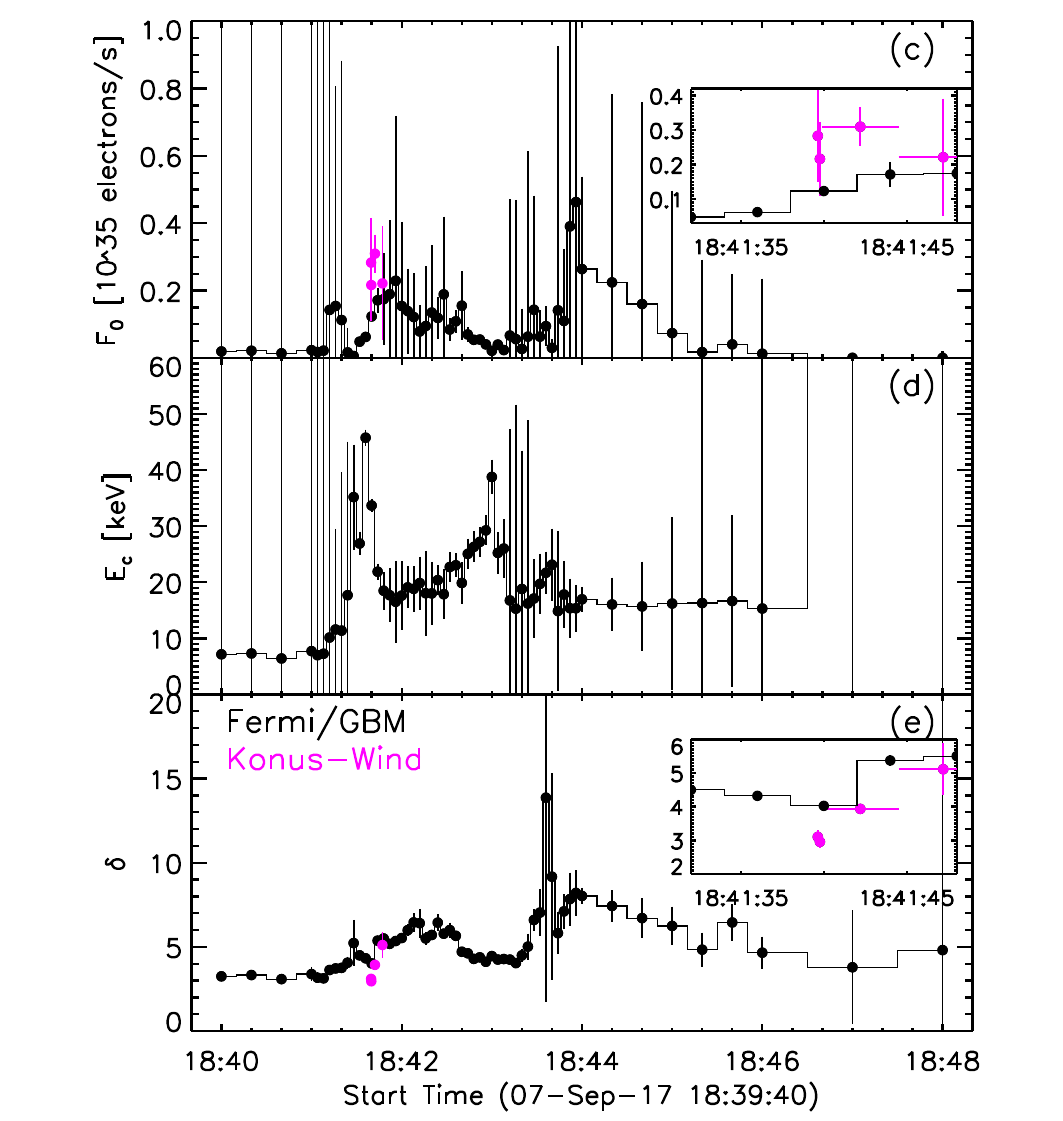}
\caption{{Time evolution of $T$ {(a)}, $EM$ {(b), and nonthermal parameters $F_0$, $E_c$, $\delta$} (c, d, e)} for the  2017-Sep-07 flare. {(a)} \fermi/GBM temperature (black) from the isothermal model for spatially unresolved full solar disk, temperature calculated from  \goes\ (red), and AIA temperature (green) from the ROI shown in Fig.\,\ref{Fig_aiaFOV}.  {(b)}  \fermi/GBM (black),  \goes\ (red), and preflare-subtracted AIA (green) emission measure, inferred as described for the upper panel. {Magenta} histogram indicates the \kw\ 18.1-76.5 keV light curve. 
(c) Total integrated electron flux $F_0$, (d) low-energy cut-off $E_c$, and
(e) spectral index $\delta$ obtained from the \fermi\ GBM (black histogram) and \kw\ ({magenta} circles) isothermal+thick target fit. 
Vertical lines indicate the range of 1$\sigma$ error on the fits. 
}
\label{Fig_EM_T}

\end{figure*}

\subsection{\fermi/GBM data} \label{S_Observations_fermi}
{Solar flare observations with \fermi/GBM in the energy range 8 keV--40 MeV provide information on properties of both thermal plasma and nonthermal electrons. For our analysis, we select CSPEC data at 128 energy channels, with 4.096\,s nominal time resolution and 1.024\,s during the burst, {provided by the} sunward-facing detector n5. 
The \fermi/GBM X-ray data with background subtracted were fitted with an isothermal (\texttt{f\_vth}) plus a collisional thick target (\texttt{f\_thick2}) model, using the OSPEX\footnote{For OSPEX documentation see 
\url{https://hesperia.gsfc.nasa.gov/rhessi3/software/spectroscopy/spectral-analysis-software/index.html}.} SSWIDL application.

These fits were applied every 20\,s from 18:40:00~UT to 18:41:00~UT, at the rise phase of the flare, every 4\,s from 18:41:00~UT to 18:44:00~UT, 
every 20\,s from 18:44:00~UT to 18:46:00~UT, and every 60\,s from 18:46:00~UT to 18:49:00~UT. The fitted intervals are highlighted by gray graduation areas in Figure~\ref{Fig_lightcurves}(a). 
The two examples of \fermi/GBM fits shown in Figure~\ref{Fig_fermi_fits} demonstrate
a transition from a nonthermal-dominated phase (middle panel) to a thermal-dominated one (right panel), a few seconds after the HXR peak at 18:41:39.9~UT. 
The time evolution of the fit parameters is shown in Figure~\ref{Fig_EM_T}: the temperature $T_{\rm GBM}$ (a), the emission measure $EM_{\rm GBM}$ (b), the total integrated electron flux $F_0$ [in units of $10^{35}$ electrons/s] (c), the low-energy
cut-off $E_c$ [keV] (d), and the spectral index $\delta$ of the electron distribution function above $E_c$ (e). The figures confirm an existence of a nonthermal component during the HXR peak. {The nonthermal electron flux shows three main peaks corresponding the three peaks of HXR emission. The third electron peak seems the largest, although the uncertainties of the electron flux are rather large, while the spectral slope is softer {and $E_c$ is smaller} here than in the main peak. }

The results of the \fermi/GBM spectral model fits are further used to quantify the thermal energy of the hottest component of the flaring plasma, as well as the nonthermal energy of accelerated electrons.
}

\subsection{\kw\ data}
{\kw\ provides spectral data for nonthermal hard X-ray and gamma-ray emission in solar flares for energies from $\sim$20 keV up to  $\gtrsim$10 MeV   \citep{2022ApJS..262...32L}. 
For our analysis, we use the \kw\ data in the trigger mode, which provide 64 energy spectra with 64\,ms accumulation time for the first four spectra, which then adaptively increases. To cross-check the \fermi/GBM fit results we consider the first six time bins after the \kw\ trigger, which cover the HXR peak.  

\begin{figure*}[!h]
\centering
\includegraphics[width=0.95\textwidth]{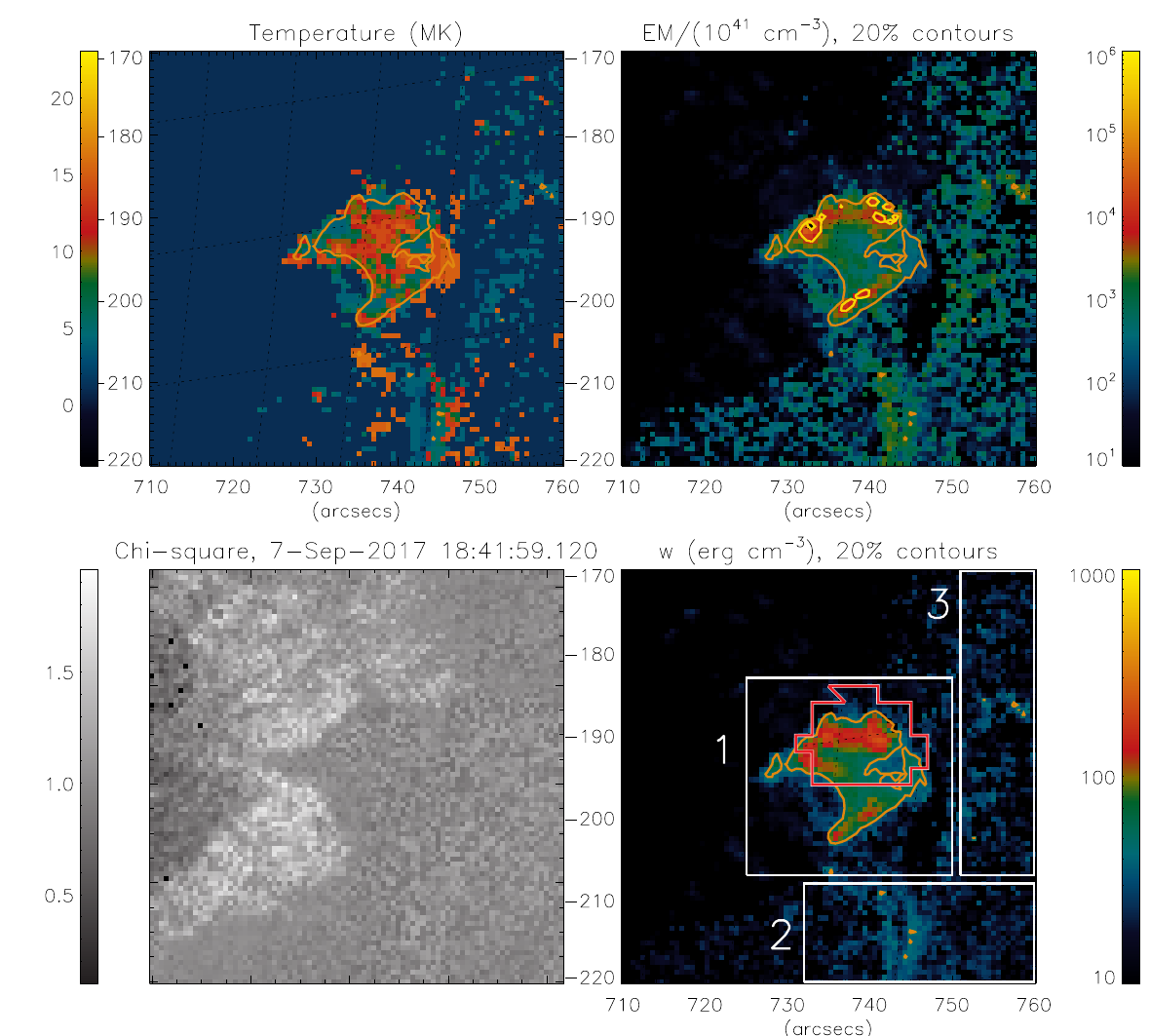} 
\caption{{Maps of plasma parameters  obtained with the regularized DEM maps based on \sdo/AIA data: 
the mean temperature map (top left), the emission measure map (top right), $\chi^2$ map (bottom left), and the thermal energy density map (bottom right) for the 10th time interval (18:41:59-18:42:11~UT). 
The boxes are labeled with the numbers, their thermal energy evolution is shown in Figure~\ref{W_boxes}.
The orange contour shows 20\% of the thermal energy density maximum, while yellow contour -- 20\%  of the emission measure peak. The mean temperature map (top left) is plotted only for those pixels where the thermal energy density exceeds 10\% of the maximum value for each time interval. {Red contour indicates the ROI shown in Figure~\ref{Fig_fit_parms}.}
On-line animation, 22 s, showing evolution of these four panels, is available.}
\label{Fig_EM_map}
}
\end{figure*}

\begin{figure*}
\centering
\includegraphics[width=0.48\textwidth]{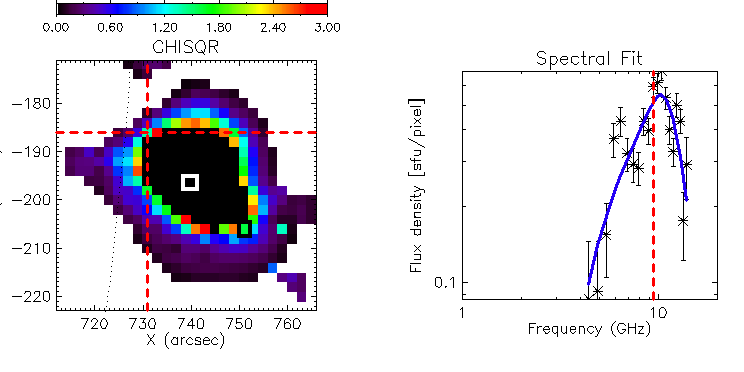}
\includegraphics[width=0.24\textwidth]{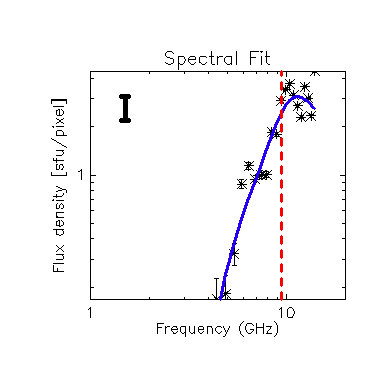}
\includegraphics[width=0.24\textwidth]{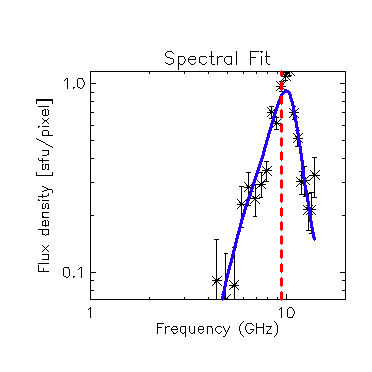}
\caption{Quality of the spectral fit. Left: $\chi^2$ map for the time frame of 18:41:39.500 UT capped at $\chi^2=3$. Other panels: examples of the spatially resolved spectra and their corresponding fits from the locations indicated by the red cursor and white and black squares, respectively. Note a relatively large scatter of the data points in the spectrum taken in the center of the image (white square; third panel), where systematic uncertainties (black vertical bar in the third panel) due to errors in the absolute flux calibration at individual frequencies are larger than the statistical error resulting in a larger $\chi^2$ in that area.
\label{Fig_fit_examples}
}
\end{figure*}

The \kw\ background-subtracted data were fitted using OSPEX, similarly to Section~\ref{S_Observations_fermi}: with an isothermal (\texttt{f\_vth}) plus a collisional thick target (\texttt{f\_thick2}) model. Because \kw\ is not sensitive to energies less than $\sim20$~keV, we use as fixed input parameters: 
emission measure $EM=6\times10^{46}$ [cm$^{-3}$], temperature $T=18.96$ [MK], and energy cut-off $E_c=18.49$ [keV] obtained from the time interval of the \fermi/GBM fit at 18:41:48-18:41:52~UT.
An example of the \kw\ fit at the HXR peak is shown in Figure~\ref{Fig_fermi_fits} left panel.
The fit results are plotted with {magenta} circles in Figure\,\ref{Fig_EM_T}c,e.
The first four 64\,ms bins, accumulated from 18:41:39.637~UT to 18:41:39.893~\,UT, were grouped onto two 128\,ms bins to improve the signal-to-noise ratio. The other two bins have a longer accumulation time of 4.6 and 5.38\,s, respectively,  from 18:41:39.893~UT to 18:41:44.501~UT, and from 18:41:44.501~UT to 18:41:49.877~UT. 
The {fit parameters during the first two short (128 ms) time intervals are different from those obtained from \fermi\ fits obtained for longer (4 s) time intervals, which might indicate sub-second variations of the nonthermal electron population.} The fit results reveal soft-hard-soft
spectral shape variability {during the main impulsive peak}, which is typical for impulsive flares with short-scale nonthermal emission \citep{2004A&A...426.1093G}.



\

}

\section{Thermal {plasma diagnostics with \sdo/AIA}} \label{aia_diagnostics}

{To account the thermal plasma of the flare, we use a combination of the six coronal \sdo/AIA  wavebands (see Section~\ref{S_Observations_aia}).
However, it should be noted that \sdo/AIA is only faintly sensitive to plasma much hotter than $\sim10$ MK and additional constraints are necessary, such as supplementing with X-ray data \citep[e.g.,][]{2015Ge&Ae..55..995M, 2019ApJ...872..204B}. Thus, AIA data can only be used to quantify the relatively cool component of the flare plasma.

To infer the emission measure (EM) [cm$^{-3}$] and temperature [K] from the ROI, we first calculate the DEM [cm$^{-5}$ K$^{-1}$] based on a regularization technique \citep{ti63, 2012A&A...539A.146H}. The errors on the AIA data (DN)  including the systematic error
have been calculated by the formula $DN_{\rm err}=(DN+(0.2\times DN)^2)^{0.5}$ (e.g., \citealt{2013ApJ...779..107B}) 
The evolution of the integral plasma parameters with the minimum preflare emission measure subtracted is given in Figure\,\ref{Fig_EM_T}(a, b): the emission measure $EM_{\rm{AIA}}$ and the mean temperature $\langle T_{\rm{AIA}} \rangle$  obtained with DEM from the ROI (see Equations (1-2), \citealt{2020ApJ...890...75M}). 
After the HXR impulsive peak, both the emission measure and temperature rise. The values of the $EM$ obtained from \fermi/GBM\ and \sdo/AIA match each other over most of the flare {evolution}, while {they diverge after roughly 18:45\,UT. Given that we do not have HXR imaging data, this may be due to X-ray emission outside our area of interest {because there maybe more plasma with a modest temperature, which} \fermi\ {starts} seeing {once the hot source has cooled down.} The temperatures {inferred from these two instruments} are noticeably different. This difference  could be due to {the fact that the} two instruments are sensitive to different temperatures and, thus, the estimates of the temperature can be biased, or the two instruments may see different sources with similar $EM$.

The spatially resolved \sdo/AIA data can provide a detailed diagnostics of the thermal plasma. 
To infer the DEM maps, the regularized inversion code \citep{2013A&A...553A..10H} has been applied to the same \sdo/AIA dataset. Here we use the methodology described in Section 2.3.2 of \citet{2020ApJ...890...75M}. The obtained DEM maps with the background DEM subtracted  are then used to calculate the emission measure maps $EM^{\rm{AIA}}_{ij}$ [cm$^{-3}$], the mean temperature maps $\langle T^{\rm{AIA}}_{ij} \rangle$ [K], and the thermal energy density $w^{\rm{AIA}}_{ij}$ [erg cm$^{-3}$] for all $(i, j)$ pixels (see Eq.(3-5),  \citealt{2020ApJ...890...75M}).
To estimate $w^{\rm{AIA}}_{ij}$, the adopted length along the line of sight (LOS) $l_{\rm depth}=d_{\rm depth}\times 0.6\times 7.25 \times 10^7$ [cm], where {0.6\arcsec\ is the AIA pixel size, $7.25 \times 10^7$ is the length  of 1\arcsec\ in cm, while}  $d_{\rm depth}=d_{\rm width}=5$ [px] has been taken as the mean loop width 
seen in the EM map for the time interval 18:42:11-18:42:23~UT.
Figure~\ref{Fig_EM_map} shows the 9th time interval (18:41:59–18:42:11~UT) of the animated version of $EM^{\rm{AIA}}$, $\langle T^{\rm{AIA}} \rangle$, and $w^{\rm{AIA}}$ maps, along with the chi-squared map (bottom left), which was less than 2 for all considered times indicative of {an} overall acceptable fitting.
The mean temperature {derived from the DEM maps, see animated Figure\,\ref{Fig_EM_map},} is around 4 MK during the HXR peak time, increases to $\sim$15 MK after it, and then slowly decreases. 
The EM and the thermal energy density sharply increase just after the nonthermal HXR peak time and then slowly decrease. 
The maximum value of the emission measure was $6\times 10^{46}$ [cm$^{-3}$]. 

}



\section{Flare diagnostics from \MW\ observations}
\subsection{Overview}

We performed the model spectral fitting of the EOVSA data with 2"$\times$2" pixel size and 1\,s time step (best time resolution available with EOVSA). 
We attempted several runs of the spectral fit using GSFIT widget \citep{2020Sci...367..278F}.
We performed an initial fitting trial over the frequency range $<15$\,GHz with six free parameters: {magnetic field strength $B$, its angle to the line of sight $\theta$, thermal number density $n_{th}$, nonthermal number density $n_{nth}$, spectral index $\delta$, and high-energy cut-off  in the electron energy distribution $E_{\max}$; see supplemental video in \cite{2022Natur.606..674F} for the dependence of the \gs\ spectrum on these parameters}. When $E_{\max}$ was well constrained, it clustered around $E_{\max} \sim2$\,MeV. We then  fixed 
$E_{\max}$ at 2\,MeV and performed several fits with five free parameters within the same ($<15$\,GHz) and  restricted ($<13.5$\,GHz) spectral ranges. The overall results of the spectral fitting are not sensitive to these choices. Here, we employ the spectral fit run with five free parameters and $f_{\max} =13.41$\,GHz.
{Figure\,\ref{Fig_fit_examples} gives an example of the $\chi^2$ map and three individual spectra with their fits. Note that systematic uncertainties (mismatches between neighboring spectra data points) exceed the statistical ones (shown as error bars) in the central area of the image.}

\subsection{Evolving parameter maps}
\label{S_ev_par_maps}


Figure\,\ref{Fig_fit_parms} shows examples of the parameter maps and the parameter evolution obtained form the spectral model fitting. All maps contain edge artifacts mostly located outside the selected EOVSA reference contour shown in black. These artifacts are known to originate from the frequency-dependent spatial resolution of the array. The fitting reveals two distinct areas (upper left and bottom right), where some of the fit parameters are different, which we interpret as a signature of two distinct flaring loops. In some of the maps, these areas are separated by a sharp line containing artifacts; likely, due to sharp changes of some physical parameters between these two areas.   

The magnetic field is remarkably similar in these two sources being within 500--600\,G. The magnetic field does not show any prominent evolution during the burst. The thermal number density is also not much different between those two areas, $n_{th}\sim10^{11}$\,cm$^{-3}$. Nonthermal electron number densities above the adopted low-energy cutoff $E_{\min}=15$\,keV are different between the two sources: it is about 10$^8$\,cm$^{-3}$ in the upper-left source, while 10$^{10}$\,cm$^{-3}$ in the bottom-right source. The spectral indices are also different such as the upper-left spectra are harder than the bottom-right ones.




These fit results show a remarkable soft-hard-soft spectral evolution during the main radio burst at both sources with the range of $\delta$ variation from 15 down to 3 and then back to 15. Interestingly, this $\delta$ variation is accompanied by apparently unexpected anti-correlation between the total number of nonthermal electrons and the radio flux, such as this number density reaches a minimum at the flare peak time. This is, likely, an artifact of the arbitrary choice of $E_{\min}=15$\,keV or of a deviation of the actual electron spectrum from the adopted here single power-law. Indeed, the \mw\ spectrum is mostly sensitive to nonthermal electrons with energies much larger than $E_{\min}=15$\,keV. Given the prominent hardening of our model spectrum at the flare rise phase, the number density of electrons above 100\,keV correlates with the \mw\ flux in agreement with the common sense{, as illustrated in Figure\,\ref{Fig_n_nth_100keV}}. 


\begin{figure*}\centering
\includegraphics[width=0.95\textwidth]{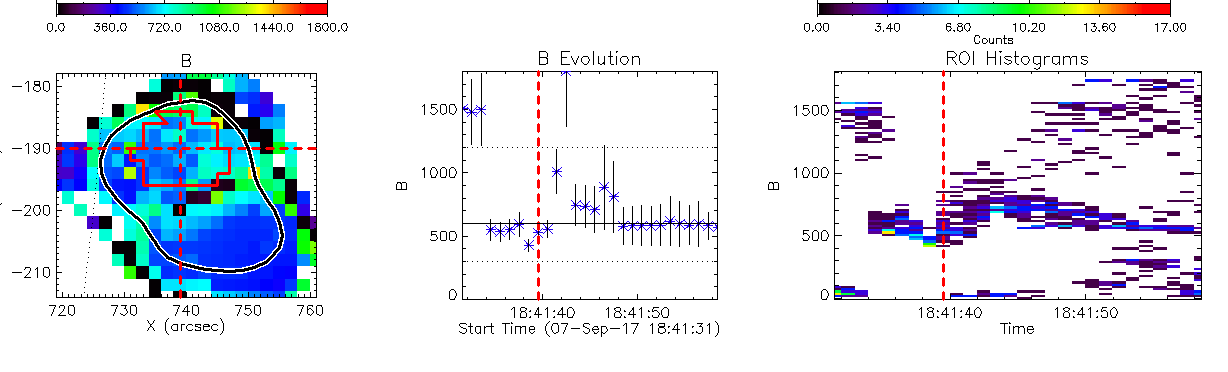}
\includegraphics[width=0.95\textwidth]{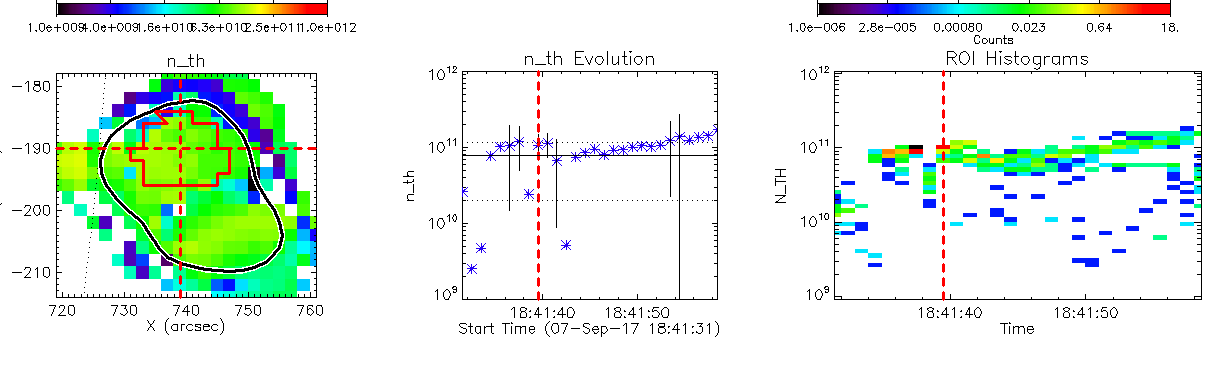}
\includegraphics[width=0.95\textwidth]{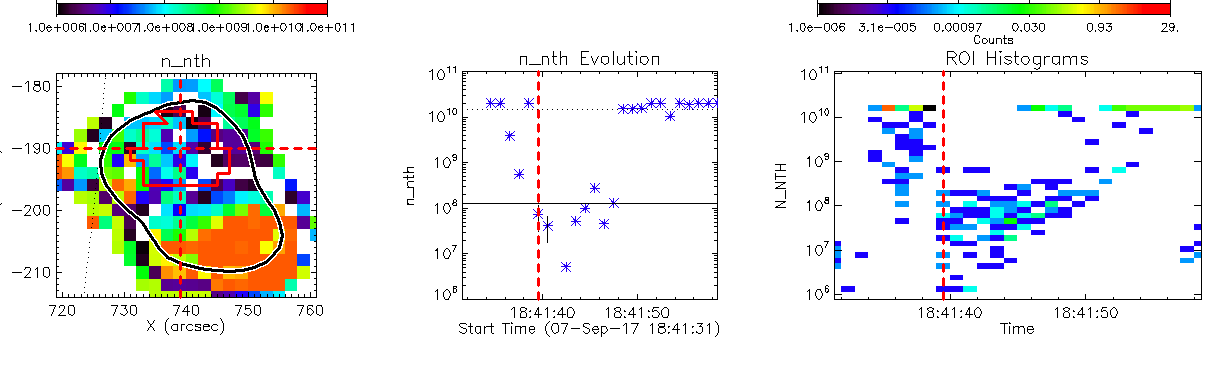}
\includegraphics[width=0.95\textwidth]
{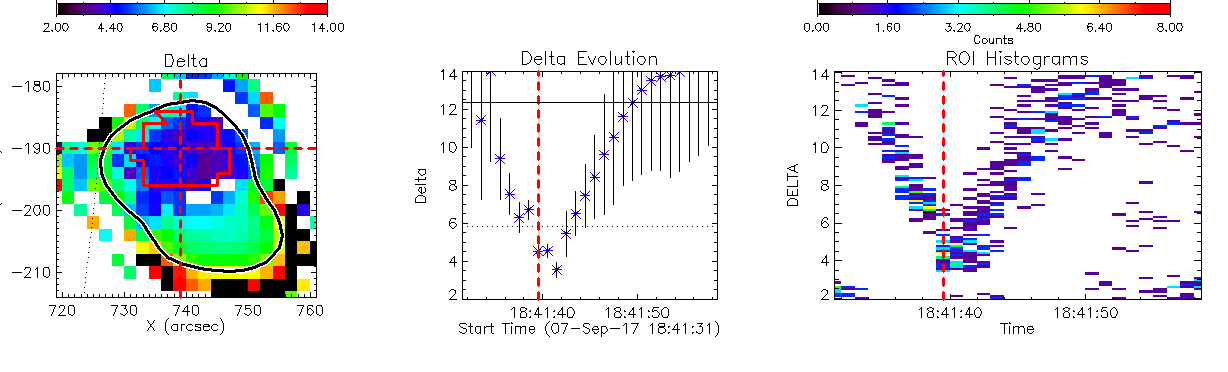}
\caption{Inferred Parameters and Their Evolution. Top Row: Magnetic field strength. Second Row: Thermal electron number density. Third Row: Non-thermal electron number density. Bottom Row: Non-thermal electron power-law index. Left Column: Parameter maps corresponding to a specific time frame (18:41:39.500 UT). Each map includes the 15\% contour (in black) of the EOVSA 10.41 GHz microwave map, the boundary defining the selected region of interest (ROI) in red, and the position of a chosen pixel within this ROI (indicated by red-dotted lines). Middle Column: Evolution of the fit parameters for the selected ROI pixel (blue symbols) along with their corresponding median values (horizontal black line). Right Column: Evolution of parameter distributions for
{all pixels of the selected ROI,}
corresponding to the same time frames as shown in the middle column plots. In both the middle and right columns, vertical dotted red lines highlight the selected time frame of the maps displayed in the left column. {The animated version, 15 s, of this figure is available, which shows evolution of the parameter maps, such as displayed in the left column of this figure, over the entire flare duration.} 
\label{Fig_fit_parms}
}
\end{figure*}

\begin{figure}\centering
\includegraphics[width=0.48\textwidth]{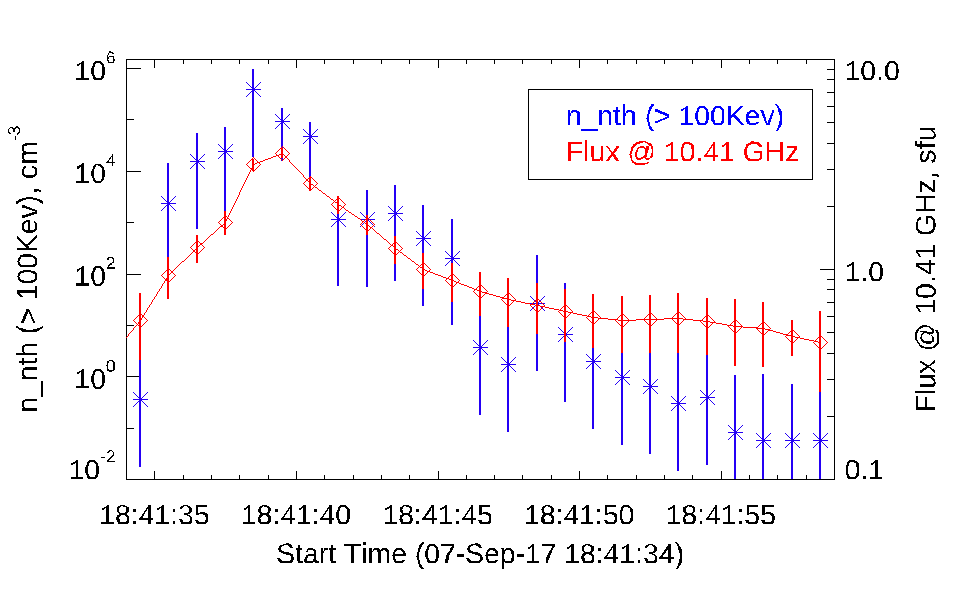}
\caption{{Evolution of the nonthermal number density above 100\,keV and fit uncertainties (blue lines and symbols) in the pixel marked by the cursor in Fig.\,\ref{Fig_fit_parms} computed from the spectral fit results shown in that figure. The EOVSA flux at 10.41 GHz measured in the same pixel is overlaid for comparison (red lines and symbols).}
\label{Fig_n_nth_100keV}
}
\end{figure}

\section{Modeling {of the flaring loops} with GX Simulator} \label{S_modelind}

We employed the automated model production pipeline \citep[AMPP,][]{nita2023gx} to create a 3D magnetic data cube based on nonlinear force-free-field (NLFFF) extrapolation constrained by an \sdo/HMI vector magnetogram taken on 18:34:41\,UT (target time was 18:40\,UT
) prior the flare. This data cube was then used to create a flare model by selecting the designated flare flux tubes to be populated  with nonthermal electron components and thermal plasma. An apparent complication in creation this model is that the magnetic field is very complex spatially at the flare site; see Figure\,\ref{Fig_connectivity}. There are many field lines that intersect in the projection {to the flare sources}; thus, creating an ambiguity in the selection of the flaring flux tubes for modeling. To double check the validity of our modeling approach, we also created a NLFFF data cube after the flare that was found to display essentially the same connectivity, which demonstrates that the magnetic field morphology did not change significantly during this flaring episode.





{Following the \mw\ diagnostics, Section\,\ref{S_ev_par_maps},}
the model we created contains two adjacent flux tubes, which originate from two northern footpoints located very close to each other. However, their connectivities are very different from each other: their southern footpoints are located far away from each other {such as their loop tops project onto two distinct \mw\ sources}. In addition, we added {a third} flux tube that has a magnetic field about 20\,G at the loop top and produces a \gs\ component with the spectral peak around 2.5\,GHz, which is used to account for the spectral flattening at low frequencies.  These flux tubes are then populated with thermal and nonthermal electrons as guided by the GSFIT results and the AIA and Fermi data {and then fine-tuned, by trials and errors, to match the \mw\ spectral and imaging data and X-ray spectral data}; an overview of the final model, Figure\,\ref{Fig_GX_model}, is given in Table\,\ref{table_model_summary_R1}. {Figure\,\ref{Fig_GX_model} displays a visualization of the 3D model by showing three involved flux tubes and spatial distributions of the nonthermal electrons inside them. It also shows synthetic-to-observed \mw\ image comparison at three different frequencies as well as observed and model \mw\ spectra at the flare peak time. The \mw\ emission from shorter loops 1 \& 2 and longer loop 3 are shown separately to demonstrate that loop 3 contribution is only significant at the lowest frequencies. Table\,\ref{table_model_summary_R1} shows some global parameters of the 3D model obtained either by direct inspection of the model geometry and numeric input or by integration over the flaring loops involved.}

\section{Comparison between the magnetic field model and spectral fitting}

To make a comparison between the magnetic field derived from the \mw\ model spectral fitting with that in the 3D model, we compute the maps of the magnetic field $B_{weighted}(x,y)$ in the model weighted with the 3D spatial distribution of nonthermal electrons $n_{nth}(x,y,z)$ by sampling the model volume over all lines of sights 
\begin{equation}
B_{weighted}(x,y) = \frac{\int\limits_0^{h_{\max}}B(x,y,z)n_{nth}(x,y,z)dz}{\int\limits_0^{h_{\max}}n_{nth}(x,y,z)dz}, \end{equation}
and then compare these maps with those obtained from the model spectral fitting; see Figure\,\ref{Fig_fit_parms}. This weighting is relevant for the comparison because the magnetic field inferred from the \mw\ spectral fitting is an averaged field over the line of sight segment occupied by the emitting nonthermal electrons. 

Figure\,\ref{Fig_B_Sampled_R1} shows the magnetic field {and, for completeness, also the nonthermal and thermal densities,} in the model volume sampled with the spatial distribution of the nonthermal electrons, where the flux tube 3 was excluded. The reason for the exclusion is that this flux tube contributes only at low frequencies, while the spectral fitting is dominated by a higher frequency portion of the spectrum; thus, the outcome of the spectral fitting is mainly relevant for comparison with flux tubes 1 and 2. This figure shows that the magnetic fields in the adjacent portions of these flux tubes are comparable to each other, $\sim500$\,G and also to the range of magnetic field values obtained from the model spectral fitting in Figure\,\ref{Fig_fit_parms}. {Two other maps show similar ranges of parameter variation as shown in Figure\,\ref{Fig_fit_parms}.} In a general case, having \gs\ emission from two different, closely located flux tube would result, due to smearing of these contributions by the finite instrumental PSF, in broadened spatially resolved spectra indicative of an inhomogeneous source, whose fitting with the uniform source cost function could be problematic. In our case, however, the ranges of the magnetic field in these two flux tubes are similar. Likely, this is why the fits are generally successful and the derived magnetic field values show rather smooth spatial behavior.


\begin{figure*}\centering
\includegraphics[width=0.49\textwidth,bb=50 50 750 750]{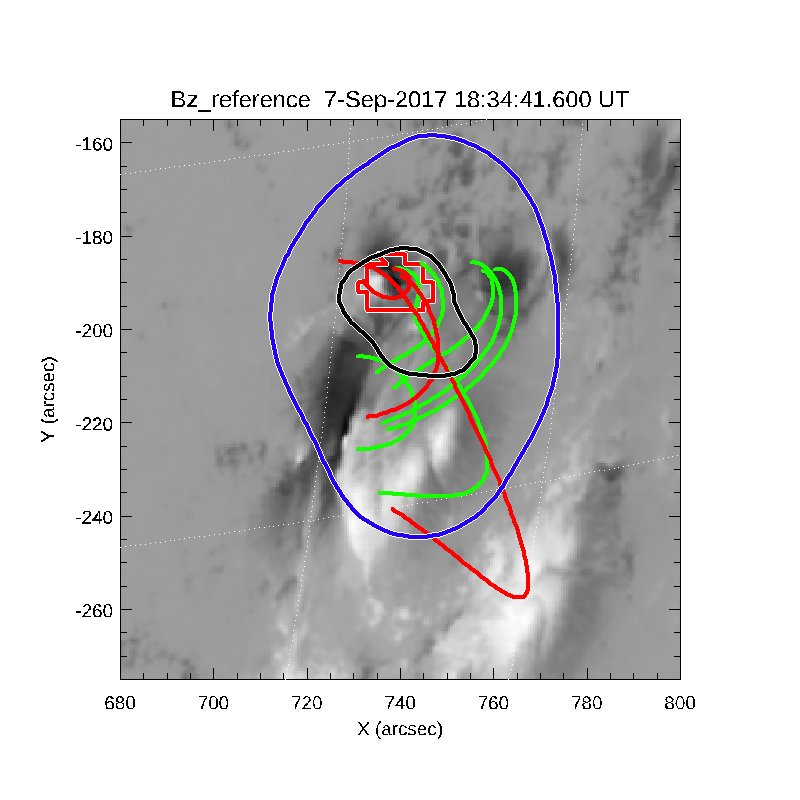}
\includegraphics[width=0.49\textwidth,bb=50 50 750 750]{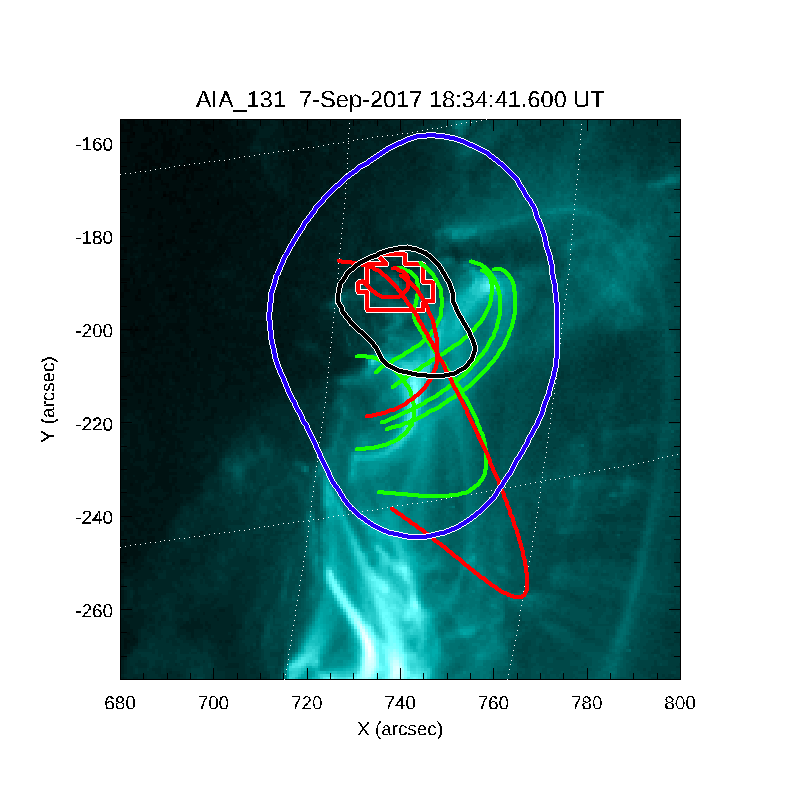}
\caption{ Visualization of 3D model connectivity. Left: three red lines show the axes of three flux tubes used for emission simulation {shown in Fig.\,\ref{Fig_GX_model} and several} additional closed field lines (green) to emphasize the complexity of the connectivity on top line-of-sight photospheric magnetic field. {We also show the ROI (red) and EOVSA 10.41\,GHz contour (black) same as in Fig.\,\ref{Fig_fit_parms} as well as an additional EOVSA 3.41\,GHz contour {(blue)}.  Right: same lines and contours are shown on top of 131\,\AA\ image. These lines} show consistency of the magnetic model with the bright EUV structures.
\label{Fig_connectivity}
}
\end{figure*}

\begin{table*}[ht]
\caption{Summary of the 3D model \hspace{1.7in}}
\begin{tabular}{l l l l l}
\hline\hline
Parameter & Symbol, units &  Loop 1 
 &  Loop       2
 &  Loop       3
\\ [0.5ex]
\hline
{\textit{Geometry}:} &  & \\
\quad Length of the Central Field Line     & $l$, Mm  &  15
 &  32
 & 126
\\
\quad Looptop magnetic field & $B$, G & 458
 & 386
 &  23
\\
\quad Footpoint 1 magnetic field & $B$, G &2422
 &2033
 & 353
\\
\quad Footpoint 2 magnetic field & $B$, G &2248
 &1940
 &1907
\\
\quad Model Volume; $\left[\int n_0 dV\right]^2/\int n_0^2 dV$ & $V$, cm$^3$ &      1.80 $\times10^{26}$
 &      1.49 $\times10^{26}$
 &      3.67 $\times10^{27}$
\\
{\textit{Thermal Plasma}:} &  & \\
\quad Total Electron Number,  $\int n_0 dV$   & $N_0$ &      5.01 $\times10^{36}$
 &      2.13 $\times10^{36}$
 &      1.27 $\times10^{36}$
\\
\quad Emission Measure,  $\int n_0^2 dV$   & $EM$, cm$^{-3}$ &      1.39 $\times10^{47}$
 &      3.05 $\times10^{46}$
 &      4.37 $\times10^{44}$
\\
\quad Mean Number Density,  $EM/N_0$   & $n_{\rm th}$, cm$^{-3}$ &      2.78 $\times10^{10}$
&        1.43 $\times10^{10}$
&        3.45 $\times10^{ 8}$
\\
\quad Temperature   & $T$, MK & 7
&24
&22
\\
\quad Thermal Energy   &  $W_{\rm th}$, erg &      1.45 $\times10^{28}$
 &      2.12 $\times10^{28}$
 &      1.15 $\times10^{28}$
\\
{\textit{Nonthermal Electrons}:} &  & \\
\quad Total Electron Number,  $\int n_{\rm b} dV$   & $N_{\rm b}$ &      8.58 $\times10^{32}$
 &      2.96 $\times10^{32}$
 &      4.73 $\times10^{34}$
\\
\quad Mean Number Density,  $\int n_{\rm b}^2 dV/N_{\rm b}$   & $n_{\rm nth}$, cm$^{-3}$ &      3.73 $\times10^{ 7}$
 &      2.55 $\times10^{ 7}$
 &      2.44 $\times10^{ 8}$
\\
\quad PWL Energy Range & $E$, MeV &     0.015 $-$ 1.0
 &     0.015 $-$ 10.
 &     0.020 $-$ 0.50
\\
\quad PWL Spectral Index & $\delta$ &      3.20
 &      3.50
 &      3.55
\\
\quad Nonthermal Energy    & $W_{\rm nth}$, erg &      3.78 $\times10^{25}$
 &      1.18 $\times10^{25}$
 &      2.49 $\times10^{27}$
\\
[1ex]
\hline
\end{tabular}
\footnote{ This model roughly matches the X-ray  spectrum at 18:41:38 --- 18:41:42 UT and perfectly at 18:41:46 UT and also complies with the AIA-derived $EM$ constraints. 
}
\label{table_model_summary_R1}
\end{table*}

\begin{figure*}\centering
\includegraphics[width=0.29\textwidth]{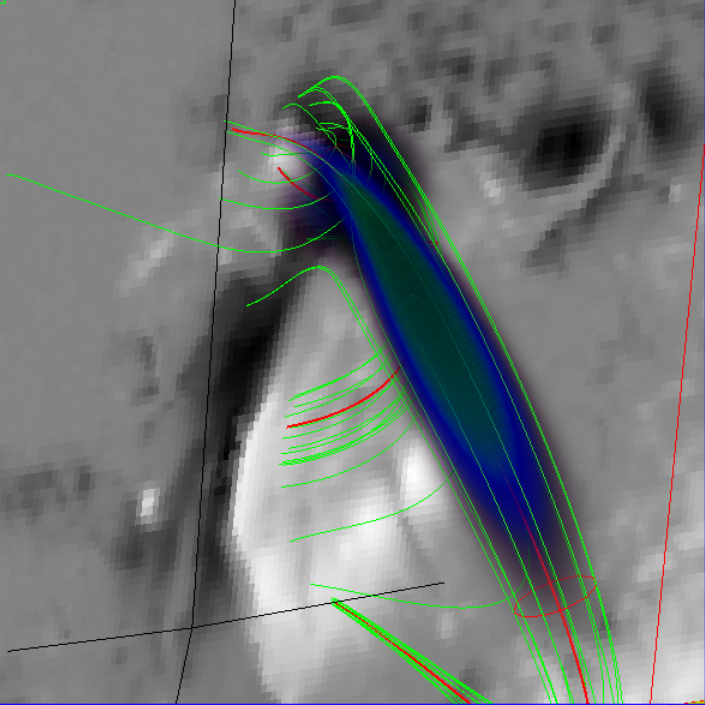}
\includegraphics[width=0.29\textwidth]{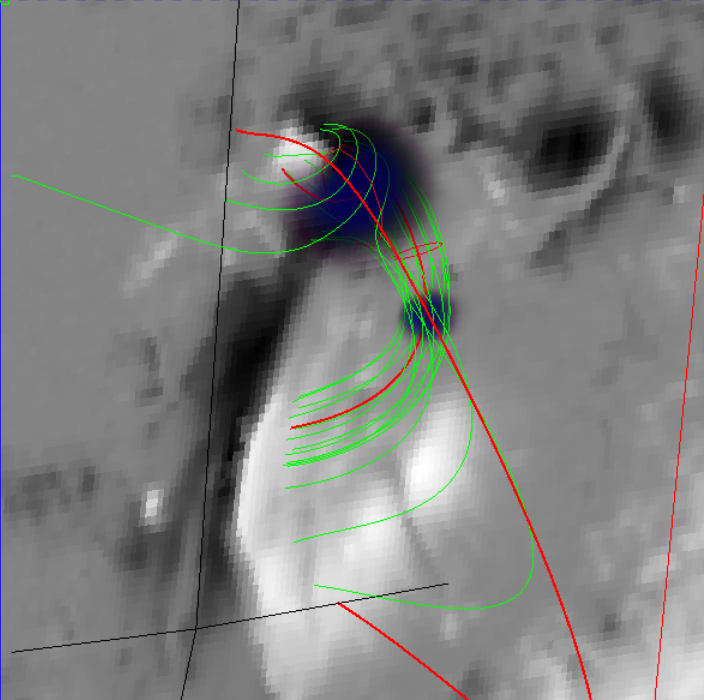}
\includegraphics[width=0.36\textwidth]{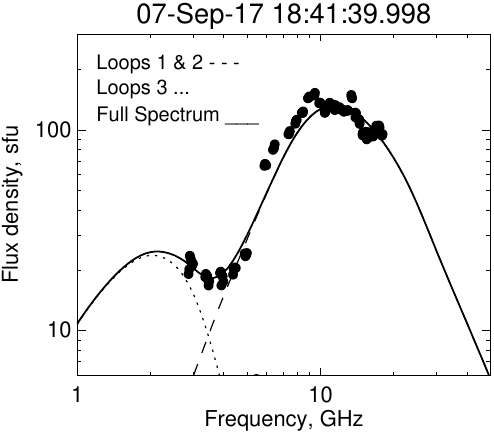}
\includegraphics[width=0.326\textwidth]{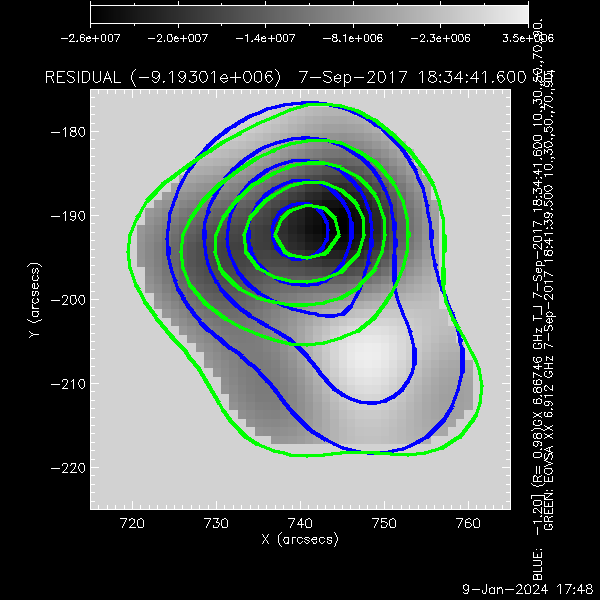}
\includegraphics[width=0.326\textwidth]{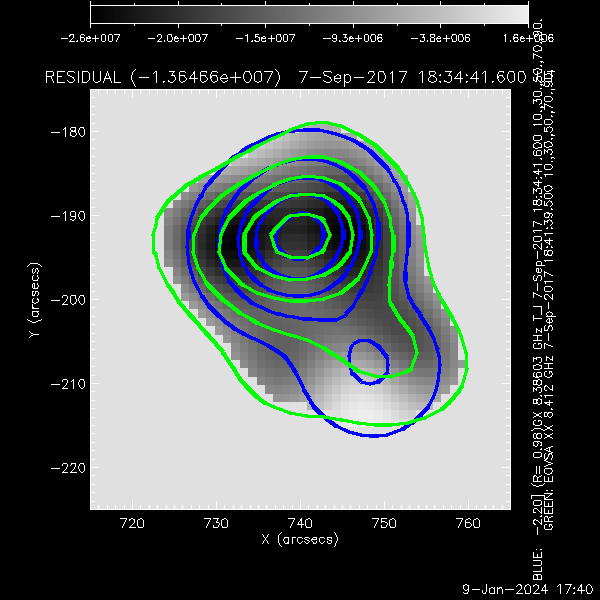}
\includegraphics[width=0.326\textwidth]{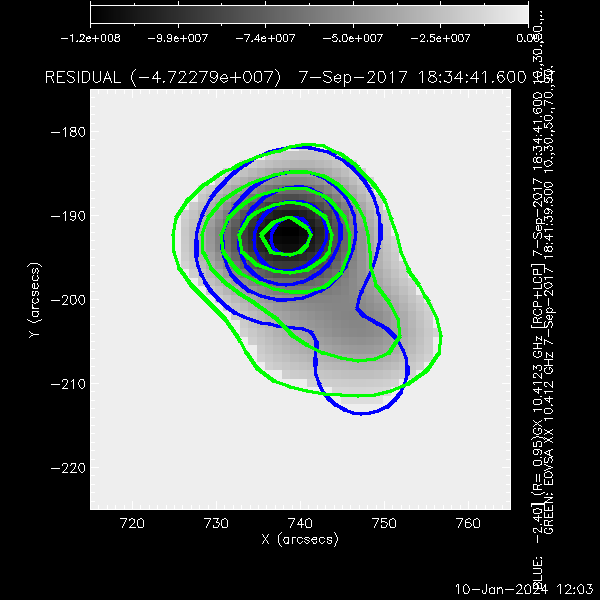}
\caption{Visualization of the 3D model. Three flux tubes are shown on top of $B_z$ magnetogram with sets of green lines, while their axes are shown with red lines. The bluish volume shows spatial distribution of the nonthermal electrons in these flux tubes. The nonthermal electron distribution in flux tube 2 is hidden under that in flux tube 3, which is excluded in the second panel (the axis of the excluded flux tube is shown as a long red line) to illustrate that. {Top right panel shows model-to-data spectral comparison: EOVSA total power data are shown with circles and the solid line shows the model spectrum integrated over the field of view shown in the previous panels; the two other lines show the contributions from flux tubes 1 \& 2 (dashed line) and 3 (dotted line).} Bottom row shows comparison between 
synthetic (blue) and observed (green) \mw\ images at three different frequencies on top of their corresponding residuals within a $50\arcsec\times50\arcsec$ FOV.
\label{Fig_GX_model}
}
\end{figure*}


\section{Energy partitions and evolution}

\begin{figure*}\centering
\includegraphics[width=0.32\textwidth]{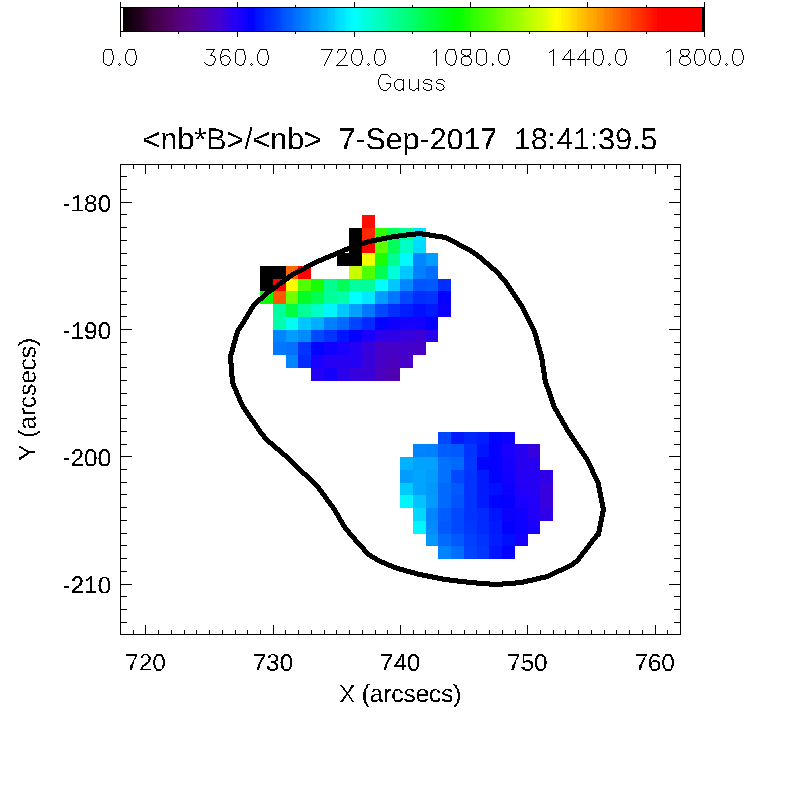}
\includegraphics[width=0.32\textwidth]{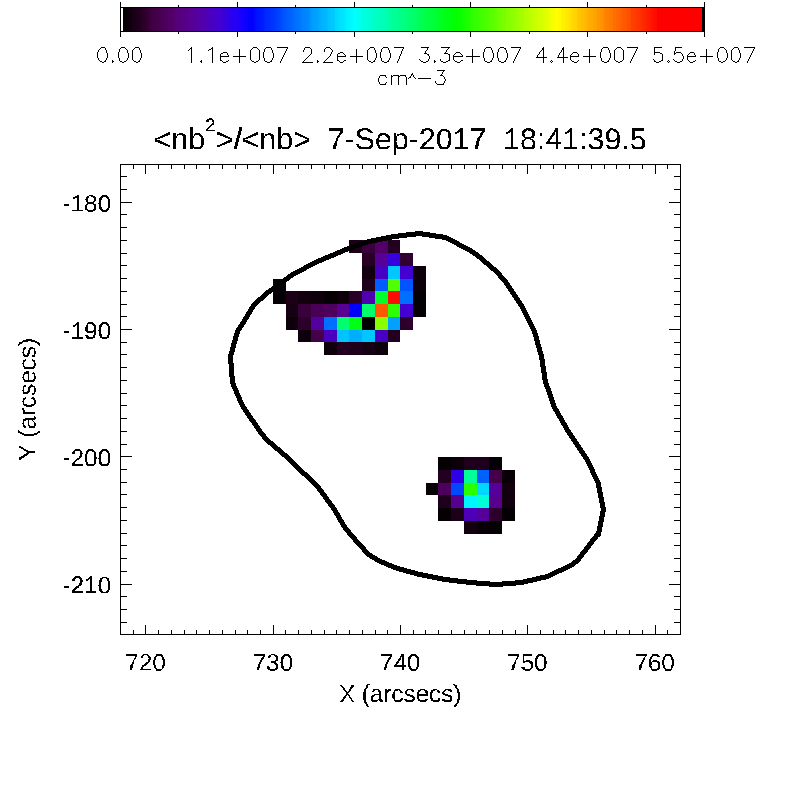}
\includegraphics[width=0.32\textwidth]{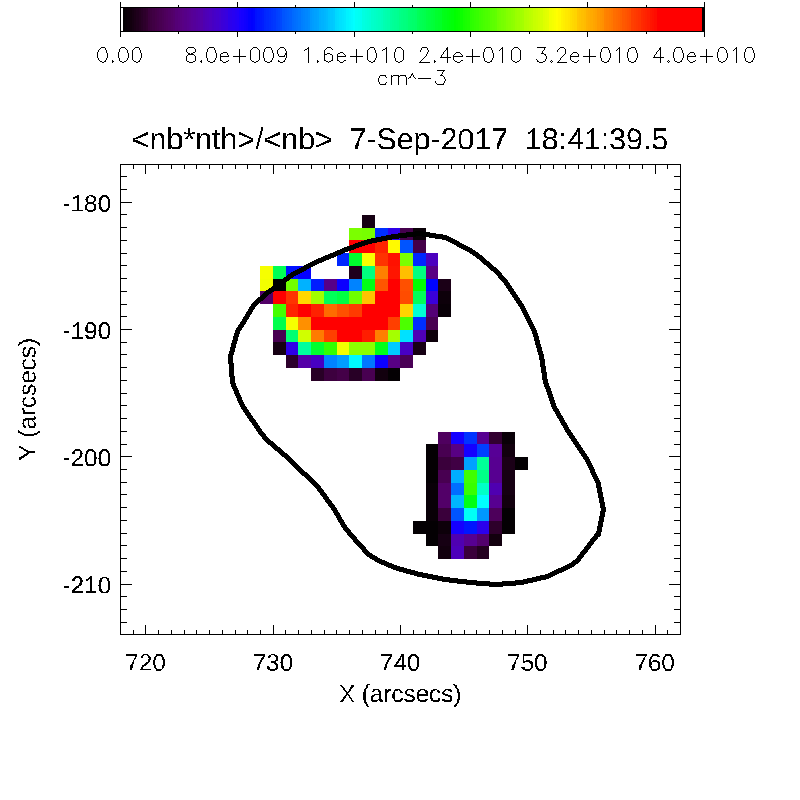}
\caption{Maps of the weighted magnetic field $B_{weighted}(x,y)${, and nonthermal and thermal number densities} for the model introduced in Figure\,\ref{Fig_GX_model} with flux tube 3 excluded shown in the same FOV as the parameter maps in Fig.\,\ref{Fig_fit_parms}. The black contour shows 15\% level of EOVSA emission at 10.41\,GHz 18:41:39.5\,UT same as in Fig.\,\ref{Fig_fit_parms}. 
\label{Fig_B_Sampled_R1}
}
\end{figure*}

\subsection{Quantification of the thermal energy with EUV data}

{The total AIA-derived thermal energy of the flare from the ROI is obtained from the thermal energy density maps $w^{\rm{AIA}}_{ij}$ described in Section \ref{aia_diagnostics}:

\begin{equation}\label{eq1}
W^{\rm{AIA}}_{\rm{th}}(t)=S_{\rm{px}}\ l_{\rm{depth}}\sum_{i=1}^{N_{\rm{px}}} \sum_{j=1}^{N_{\rm{px}}} w^{\rm{AIA}}_{ij}(t) \; [\rm{erg}].
\end{equation}
The evolution of the thermal energy $W^{\rm{AIA}}_{\rm{th}}$ with subtracted minimum preflare value of $W_{\rm th,min}^{\rm AIA} = 1.78 \times 10^{28}$ [erg] from the non-flaring pixels in the ROI is shown in Figures \ref{W_boxes} and \ref{W_final}.

The animated Figure \ref{Fig_EM_map} demonstrates that, beside the evolution of the main flaring source (in box \#1), there is some dynamics in the bottom (\#2) and right (\#3) boxes  (outlined and numbered in the Figure). It is unclear whether this dynamics is related to the flare, or  independent. 
To estimate the contribution of all boxes to the thermal energy, we show them together with the contribution $W^{\rm{AIA}}_{\rm{th}}$ from the ROI in Figure \ref{W_boxes}. 
In all cases, the minimum preflare energy from the corresponding box or ROI was 
subtracted.
Figure \ref{W_boxes} shows that the contribution of the boxes \#2 and \#3 is small compared to the flaring box \#1; however, they show a slight increase with time that correlates with the behavior of box \#1. 
This may indicate that, although the main energy release occurs in box \#1, the flaring process expands to a larger area. Thus, to take these processes into account, further we use the value of the thermal energy $W^{\rm{AIA}}_{\rm{th}}$ from the ROI (see Figure \ref{W_final}) which reaches a peak value of about $5.3\times 10^{28}$ [erg] during {the} 18:43:47-18:43:59~UT {interval}.
}

\begin{figure}\centering
\includegraphics[width=0.49\textwidth]
{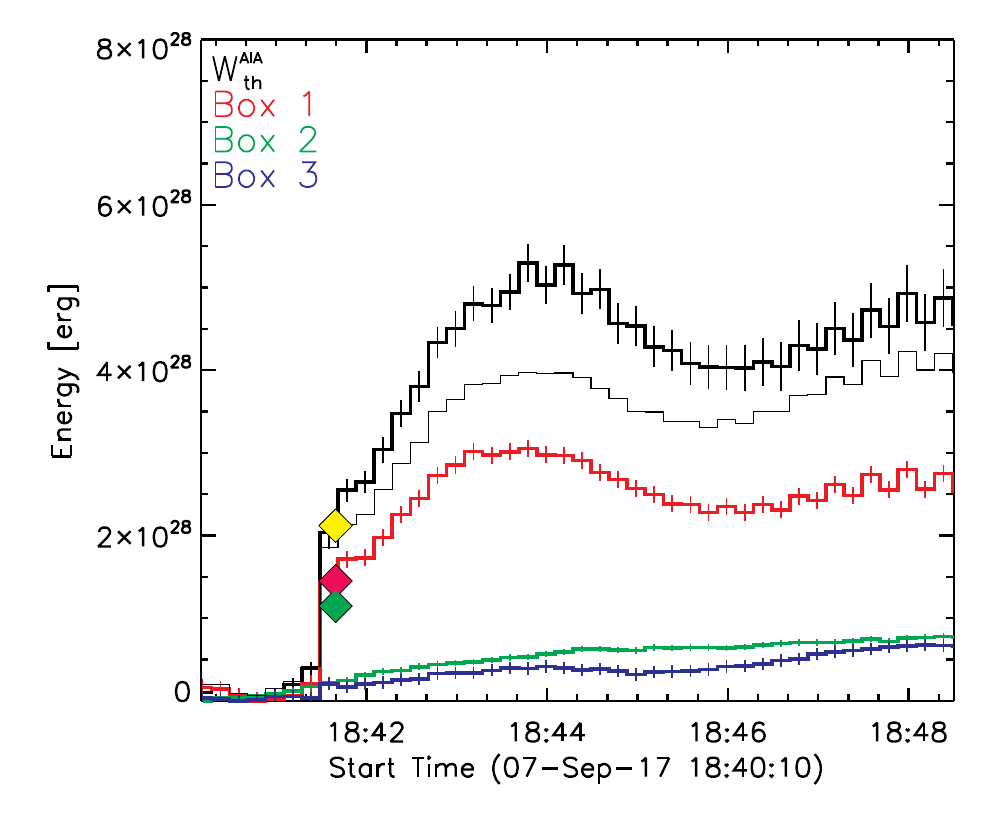}
\caption{{Evolution of the thermal energy $W_{\rm th}^{\rm AIA}$ computed from the DEM maps inside the ROI shown in Figure~\ref{Fig_aiaFOV} (black line) and 
three boxes (red, green, and blue lines) outlined by numbers in Figure~\ref{Fig_EM_map}. The sum of three boxes is shown in thin black line.
The scarlet, yellow, and green diamonds indicate the model values of thermal energies in
Loops 1, 2, and 3, respectively; see Table~\ref{table_model_summary_R1}.
}    
\label{W_boxes}
}
\end{figure}

\begin{figure}\centering
\includegraphics[width=0.49\textwidth]
{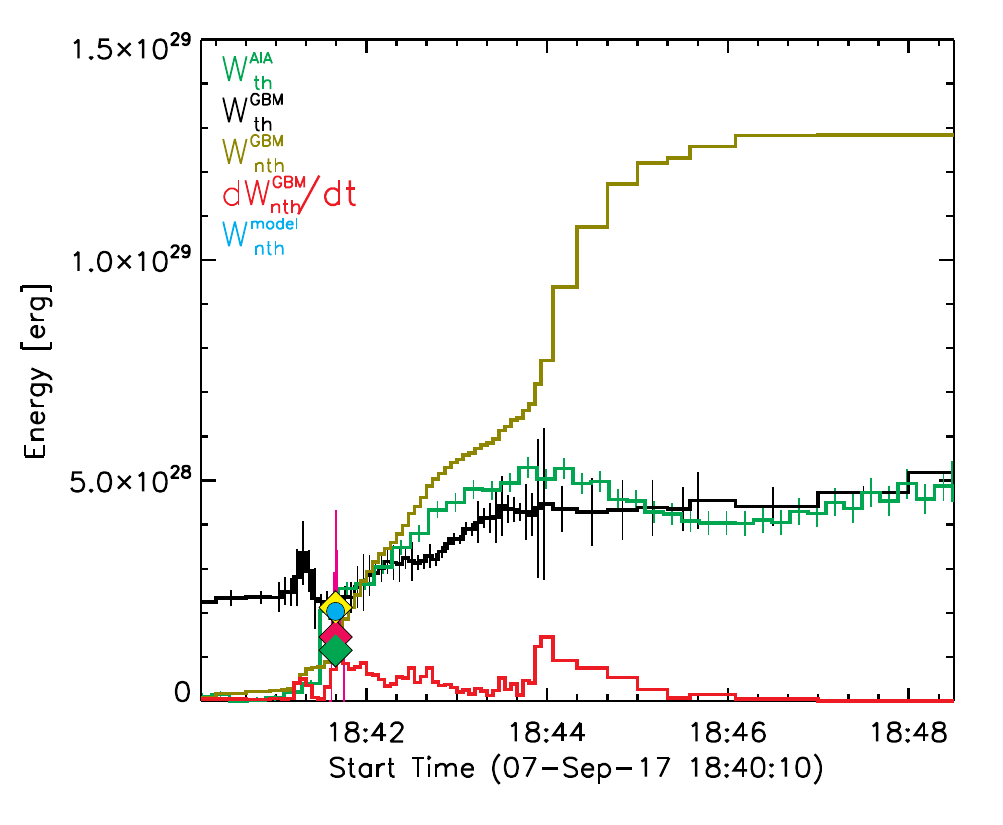}
\caption{{Evolution of energy components in the 2017-Sep-07 flare. 
The thermal energy $W_{\rm th}^{AIA}$ computed from the ROI of the AIA DEM maps is shown in green. 
The thermal energy $W_{\rm th}^{\rm GBM}$ deduced from the thermal part of the
\fermi/GBM fits assuming {flux tube 2} volume in the 3D model is shown in black. 
The cumulative nonthermal energy deposition $W_{\rm nth}^{\rm GBM}$ obtained from the nonthermal part of the \fermi/GBM fits is shown with the dark yellow histogram, while the red histogram shows the rate of \fermi/GBM nonthermal energy deposition $dW_{\rm nth}^{\rm GBM}/dt$ (arb. units). The total nonthermal energy input inferred from the three loops of the 3D model is shown {by a light-blue circle.} 
{Magenta}  {curve shows} the \kw\ 18.1-76.5 keV light curve. 
The scarlet, yellow, and green diamonds indicate the model values of thermal energies in
Loops 1, 2, and 3, respectively; see Table~\ref{table_model_summary_R1}. 
}    
\label{W_final}
}
\end{figure}

\subsection{Quantification of the thermal energy with X-ray data}

{The \fermi/GBM spectral model fits can be used to quantify the
thermal energy of the hot component ($\gtrsim$20\,MK) of the flaring plasma in a similar way to the estimate reported by \citet{2020ApJ...890...75M}.
To account for the thermal energy detected by \fermi/GBM, we employ the emission measure and temperature obtained from the \fermi/GBM fits (see Section \ref{S_Observations_fermi} and Figure \ref{Fig_EM_T}):
\begin{equation}\label{eq2}
W^{\rm{GBM}}_{\rm{th}}=3 k_B T_{\rm{GBM}} \sqrt{EM_{\rm{GBM}}\times V} \; [\rm{erg}],
\end{equation}
where $V$ is the volume of the corresponding thermal source. 
Here we cannot estimate the hot loop volume from the data directly, due to the lack of X-ray imaging data. Instead, we rely on the volume of the hottest loop, {$V=1.49\times 10^{26}$ [cm$^{3}$] (Loop 2; see Table\,\ref{table_model_summary_R1})}, determined from the 3D model devised in Section \ref{S_modelind}.
The evolution of the \fermi/GBM-derived thermal energy $W^{\rm{GBM}}_{\rm{th}}$ is shown in black in Figure \ref{W_final}. The time history of the estimated thermal energy has a similar shape to the nonthermal energy deposition, {which turns sufficient to account the observed plasma heating. We note, however, that this conclusion is based on}  our estimate of the X-ray source volume {from the 3D model rather than from imaging. 
} 
Indeed, given the lack of X-ray imaging data, we might have incorrectly ascribed the hottest plasma to  loop 2; it is possible that another, smaller {or bigger} loop with proportionally smaller{/larger} volume, is in fact the main contributor to the thermal X-ray emission. In this case, the estimate of the thermal energy would be proportionally smaller{/larger} than that shown in the Figure.

\subsection{Quantification of the nonthermal energy with X-ray and \mw\ data and 3D modeling}

{
The nonthermal energy  $W_{\rm nth}$  deposited in the flaring volume during  18:40:00-18:49:00~UT versus time was computed as a cumulative sum using the parameters from the \fermi/GBM fits:

\begin{equation}\label{eq5}
W^{\rm{GBM}}_{\rm{nth}}=\int\limits_{- \infty}^t F_0 E_c \frac{\delta-1}{\delta-2} dt \; [\rm{erg}],
\end{equation}
where $F_0$, $E_c$, and $\delta$ are the thick target parameters shown in Figure\,\ref{Fig_EM_T}c-e (see Section \ref{S_Observations_fermi}).  
$W^{\rm{GBM}}_{\rm{nth}}$ is shown in dark yellow in Figure \ref{W_final}.

We compare this energy deposition with the one estimated using the data-validated 3D model following the approach outlined by \citep{2021ApJ...913...97F}. Specifically, using the nonthermal energy estimated at a selected time frame, see Table\,\ref{table_model_summary_R1}, we can estimate the total nonthermal energy deposition during the main flare peak duration $\tau\approx8$\,s if we know the escape time $\tau_{esc}$ of the nonthermal electrons from the source. 
\citet{2021ApJ...913...97F} estimated the escape time as the time delay between the HXR and \mw\ light curves. In our case, there is no measurable time delay within the available EOVSA time resolution of 1\,s. Therefore, we can only estimate an upper bound of the escape time to be $\tau_{esc}< 1$\,s and, thus, the lower bound of the nonthermal energy deposition $W_{\rm nth}>2.4\times10^{28}$\,erg, which is consistent with the $W^{\rm{GBM}}_{\rm{nth}}$ estimated above{, see Figure\,\ref{W_final}}.


}

\vspace{-0.2cm}
\section{Discussion}

\subsection{Summary of main findings}

Here we performed a multi-instrument and multi-wavelength study of a 2017-Sep-07 ``cold'' solar flare. The uniqueness of this flare compared with other cold solar flares reported so far is the  \mw\ imaging spectroscopy available from EOVSA. This new data set permitted us to obtain evolving parameter maps of physical parameters, including the magnetic field in the coronal portion of the solar flare, and, thus, much better constrain its magnetic structure and plasma environment. 

We found that neither the magnetic field nor the plasma density show prominent evolution during the course of the flare, unlike the powerful 2017-Sep-10 eruptive flare produced in the same active region \citep{2020Sci...367..278F}. On the contrary, the nonthermal electron population shows a prominent soft-hard-soft spectral evolution with much stronger variation of the spectral index than typically reported from the X-ray diagnostics.

We estimated the thermal energy evolution constrained by the EUV and X-ray data, as well as nonthermal energy deposition constrained with the X-ray and \mw\ data and 3D modeling. We found that the thermal energies are highly correlated with the nonthermal energy deposition. This implies that the observed thermal energy signatures are the response to the deposition of the nonthermal energy. We conclude that there was no direct plasma heating in this flare and that the entire thermal emission was due to the plasma's response to the nonthermal energy deposition.

\subsection{Comparison with other cold flares} 

Let us compare the 2017-Sep-07 cold flare with previously studied cases. A key question is how much and where exactly the free magnetic energy has been released to drive the flare. In some of the previous studies, this was evaluated using 3D magnetic extrapolations before and after the flare; for example, \cite{2021ApJ...913...97F} did not find any measurable change in the magnetic energy due to the flare within uncertainties of the models. In our case, we directly measured the coronal magnetic field in the flaring volume. Although some variations are visible in the evolving magnetic field maps, the analysis does not reveal any systematic or statistically significant trends. The uncertainties of the magnetic field in Figure\,\ref{Fig_fit_parms} are about 100\,G or larger. The ROI shown in that figure inscribes about 50 pixels, which corresponds to the source volume of $V\approx6\times10^{26}$\,cm$^{-3}$. Therefore the uncertainty in the magnetic energy within the ROI is $\Delta W_B \sim 2.5\times10^{29}$\,erg, which is one order of magnitude larger than the energy needed to drive this flare. This is consistent with the lack of detectable decrease of the magnetic field in this event, in contrast with large flares \citep{2020Sci...367..278F}.

The morphology of this flare is overall similar to other cold flares, as it includes three distinct, presumably interacting, flaring loops. The energy partitions between these loops in the developed model are summarized in Table\,\ref{table_model_summary_R1}. Interesting, that most of the energy belongs to flux tube 3, while the dominant contributors to the \mw\ emission are loops 1 and 2.

The nonthermal energy deposition is sufficient for the thermal energy quantified by the AIA {and \fermi} data; Figure\,\ref{W_final}.  We conclude that the scenario in which the entire thermal response in this flare is driven by nonthermal energy deposition is likely, similar to other cold flares.


\acknowledgements
This work was partly supported 
by NSF grants RISE-2324724 and  
AGS-2425102  
and NASA grant 
80NSSC23K0090 
to New Jersey Institute of Technology, 
GM (X-ray and EUV analysis) was supported by RSF grant No. 20-72-10158.
{We are thankful to Alexandra Lysenko for fruitful discussions of the \kw\ data.}

\bibliography{all_issi_references,2017sep10,fleishman}

\end{document}